\documentstyle[12pt]{article}
\def\hybrid{\topmargin 0pt      \oddsidemargin 0pt
        \headheight 0pt \headsep 0pt
        \textwidth 16.5cm
        \textheight 23cm
        \voffset=-0.2cm
        \hoffset=0.4cm
        \marginparwidth 0.0in
        \parskip 5pt plus 1pt   \jot = 1.5ex}
\catcode`\@=11
\def\marginnote#1{}

\newcount\hour
\newcount\minute
\newtoks\amorpm
\hour=\time\divide\hour by60
\minute=\time{\multiply\hour by60 \global\advance\minute by-\hour}
\edef\standardtime{{\ifnum\hour<12 \global\amorpm={am}%
        \else\global\amorpm={pm}\advance\hour by-12 \fi
        \ifnum\hour=0 \hour=12 \fi
        \number\hour:\ifnum\minute<10 0\fi\number\minute\the\amorpm}}
\edef\militarytime{\number\hour:\ifnum\minute<10 0\fi\number\minute}

\def\draftlabel#1{{\@bsphack\if@filesw {\let\thepage\relax
   \xdef\@gtempa{\write\@auxout{\string
      \newlabel{#1}{{\@currentlabel}{\thepage}}}}}\@gtempa
   \if@nobreak \ifvmode\nobreak\fi\fi\fi\@esphack}
        \gdef\@eqnlabel{#1}}
\def\@eqnlabel{}
\def\@vacuum{}
\def\draftmarginnote#1{\marginpar{\raggedright\scriptsize\tt#1}}

\def\draft{\oddsidemargin -0.1truein
        \def\@oddfoot{\sl preliminary draft \hfil
        \rm\thepage\hfil\sl\today\quad\militarytime}
        \let\@evenfoot\@oddfoot \overfullrule 3pt
        \let\label=\draftlabel
        \let\marginnote=\draftmarginnote
   \def\@eqnnum{{\rm (\theequation)}\rlap{\kern\marginparsep\tt\@eqnlabel}%
\global\let\@eqnlabel\@vacuum}  }

\font\teneuf=eufm10  scaled  1\@ptsize00 
\font\seveneuf=eufm7 scaled  1\@ptsize00 
\font\fiveeuf=eufm5  scaled  1\@ptsize00 

\newfam\euffam
\textfont\euffam=\teneuf  \scriptfont\euffam=\seveneuf
  \scriptscriptfont\euffam=\fiveeuf

\def\hexnumber@#1{\ifnum#1<10 \number#1\else
 \ifnum#1=10 A\else\ifnum#1=11 B\else\ifnum#1=12 C\else
 \ifnum#1=13 D\else\ifnum#1=14 E\else\ifnum#1=15 F\fi\fi\fi\fi\fi\fi\fi}

\def\got{\ifmmode\let\next\got@\else
 \def\next{\errmessage{Use \string\got\space only in math mode}}\fi\next}
\def\got@#1{{\got@@{#1}}}
\def\got@@#1{\fam\euffam#1}

\newfont{\lgot}{eufm10 scaled 1440}%

\newfont{\Bbb}{msbm10 scaled 1\@ptsize00}
\newcommand{\CC}{\mbox{\Bbb C}}

\newcommand{\ZZ}{\mbox{\Bbb Z}}
\newfont{\Bbbb}{msbm7 scaled 1\@ptsize00}
\newcommand{\z}{\mbox{\Bbbb Z}}

\font\sevenmsa=msam6 
\newfam\msafam
\textfont\msafam=\sevenmsa
\def\hexnumber@#1{\ifnum#1<10 \number#1\else
\ifnum#1=10 A\else\ifnum#1=11 B\else\ifnum#1=12 C\else
\ifnum#1=13 D\else\ifnum#1=14 E\else\ifnum#1=15 F\fi\fi\fi\fi\fi\fi\fi}
\def\msa@{\hexnumber@\msafam}
\mathchardef\blacktriangleright="3\msa@49
\mathchardef\blacktriangleleft="3\msa@4A
\mathchardef\trianglerighteq="3\msa@44
\mathchardef\trianglelefteq="3\msa@45

\newdimen\linethick  \linethick=0.4pt
\newdimen\hboxitspace    \hboxitspace=5pt
\newdimen\vboxitspace    \vboxitspace=5pt

\def\fr#1{%
\beq\new
\vcenter{
\hrule height\linethick
           \hbox{\vrule width\linethick
                 \kern\hboxitspace
                 \vbox{\kern\vboxitspace
                       \hbox{$\begin{array}{c}\displaystyle#1
          \end{array}$}%
                       \kern\vboxitspace}%
                 \kern\hboxitspace
                 \vrule width\linethick}%
           \hrule height\linethick}%
\eeq}

\newdimen\Squaresize \Squaresize=14pt
\newdimen\Thickness \Thickness=0.5pt

\def\Square#1{\hbox{\vrule width \Thickness
   \vbox to \Squaresize{\hrule height \Thickness\vss
      \hbox to \Squaresize{\hss#1\hss}
   \vss\hrule height\Thickness}
\unskip\vrule width \Thickness}
\kern-\Thickness}

\def\Vsquare#1{\vbox{\Square{$#1$}}\kern-\Thickness}

\def\numberbysection{\@addtoreset{equation}{section}
        \def\theequation{\thesection.\arabic{equation}}}
\numberbysection

\renewcommand{\theequation}{\thesection.\arabic{equation}}
\newcommand{\l@qq}[2]{\addvspace{2em}
 \hbox to\textwidth{\hspace{1em}\bf #1 \dotfill #2}}

\newcounter{app}

\def\app{\setcounter{equation}{0}
\def\theequation{\Alph{app}.\arabic{equation}}\par
   \addvspace{4ex}
   \@afterindentfalse
  \secdef\@app\@dapp}

\newcommand\@app{\@startsection {app}{1}{0ex}%
                                   {-3.5ex \@plus -1ex \@minus -.2ex}%
                                   {2.3ex \@plus.2ex}%
                                   {\normalfont\Large\bf}}
\def\@dapp#1{%
{\parindent \z@ \raggedright  \bf #1}\par\nobreak}
\def\l@app#1#2{\ifnum \c@tocdepth >\z@
    \addpenalty\@secpenalty
    \addvspace{1.0em \@plus\p@}%
    \setlength\@tempdima{8em}%
    \begingroup
      \parindent \z@ \rightskip \@pnumwidth
      \parfillskip -\@pnumwidth
      \leavevmode \bfseries
      \advance\leftskip\@tempdima
      \hskip -\leftskip
      #1\nobreak\hfil \nobreak\hb@xt@\@pnumwidth{\hss #2}\par
    \endgroup\fi}
\newcounter{sapp}[app]

\def\sapp{\def\theequation{\Alph{app}.\arabic{equation}}
\par
\@afterindentfalse
  \secdef\@sapp\@dsapp}
\newcommand{\@sapp}{\@startsection{sapp}{2}{\z@}%
                                     {-3.25ex\@plus -1ex \@minus -.2ex}%
                                     {1.5ex \@plus .2ex}%
                                     {\normalfont\large\bfseries}}

\def\@dsapp#1{%
{\parindent \z@ \raggedright  \bf #1
}\par\nobreak}
\newcommand{\l@sapp}{\@dottedtocline{2}{1.5em}{2.3em}}

\def\titlepage{\@restonecolfalse\if@twocolumn\@restonecoltrue\onecolumn
     \else \newpage \fi \thispagestyle{empty}\c@page\z@
        \def\thefootnote{\fnsymbol{footnote}} }

\def\endtitlepage{\if@restonecol\twocolumn \else  \fi
        \def\thefootnote{\arabic{footnote}}
        \setcounter{footnote}{0}}  
\relax

\hybrid

\newtoks\@stequation

\def\subequations{\refstepcounter{equation}%
  \edef\@savedequation{\the\c@equation}%
  \@stequation=\expandafter{\theequation}
  \edef\@savedtheequation{\the\@stequation}
  \edef\oldtheequation{\theequation}%
  \setcounter{equation}{0}%
  \def\theequation{\oldtheequation\alph{equation}}}

\def\endsubequations{%
  \setcounter{equation}{\@savedequation}%
  \@stequation=\expandafter{\@savedtheequation}%
  \edef\theequation{\the\@stequation}%
  \global\@ignoretrue}

\makeatletter
\newdimen\normalarrayskip              
\newdimen\minarrayskip                 
\normalarrayskip\baselineskip
\minarrayskip\jot
\newif\ifold             \oldtrue            \def\new{\oldfalse}
\def\arraymode{\ifold\relax\else\displaystyle\fi} 
\def\eqnumphantom{\phantom{(\theequation)}}     
\def\@arrayskip{\ifold\baselineskip\z@\lineskip\z@
     \else
     \baselineskip\minarrayskip\lineskip1\baselineskip\fi}

\def\@arrayclassz{\ifcase \@lastchclass \@acolampacol \or
\@ampacol \or \or \or \@addamp \or
   \@acolampacol \or \@firstampfalse \@acol \fi
\edef\@preamble{\@preamble
  \ifcase \@chnum
     \hfil$\relax\arraymode\@sharp$\hfil
     \or $\relax\arraymode\@sharp$\hfil
     \or \hfil$\relax\arraymode\@sharp$\fi}}

\def\@array[#1]#2{\setbox\@arstrutbox=\hbox{\vrule
     height\arraystretch \ht\strutbox
     depth\arraystretch \dp\strutbox
     width\z@}\@mkpream{#2}\edef\@preamble{\halign \noexpand\@halignto
\bgroup \tabskip\z@ \@arstrut \@preamble \tabskip\z@ \cr}%
\let\@startpbox\@@startpbox \let\@endpbox\@@endpbox
  \if #1t\vtop \else \if#1b\vbox \else \vcenter \fi\fi
  \bgroup \let\par\relax
  \let\@sharp##\let\protect\relax
  \@arrayskip\@preamble}
\def\eqnarray{\stepcounter{equation}%
              \let\@currentlabel=\theequation
              \global\@eqnswtrue
              \global\@eqcnt\z@
              \tabskip\@centering                      
              \let\\=\@eqncr
              $$%
            \halign to \displaywidth  \bgroup
             \eqnumphantom \@eqnsel
      \hskip\@centering                               
    $\displaystyle  \tabskip\z@ {##}$%
    &\global\@eqcnt\@ne \hskip 2\arraycolsep
         $ \displaystyle  \arraymode{##}$\hfil
    &\global\@eqcnt\tw@ \hskip 2\arraycolsep
         $\displaystyle\tabskip\z@{##}$\hfil
         \tabskip\@centering
    &{##}\tabskip\z@\cr}
\makeatother
\newtheorem{de}{Definition}[section]        

\newtheorem{ex}{Example}[section]

\newtheorem{rem}{Remark}[section]

\def\bea{\begin{eqnarray}}
\def\eea{\end{eqnarray}}

\def\beq{\begin{equation}}
\def\eeq{\end{equation}}
\def\be{\beq\new\begin{array}{c}}
\def\ee{\end{array}\eeq}
\def\bse{\begin{subequations}}                
\def\ese{\end{subequations}}                 %

\def\al{\alpha}

\def\ve{\varepsilon}

\def\<{\langle}
\def\>{\rangle}

\def\g{{\got g}}
\def\hg{\hat{\got g}}

\def\h{{\got h}}
\def\gli{\got{gl}(\infty)}

\def\hsl{\widehat{\got{sl}}}

\def\gl{\got{gl}}
\def\hsln{\widehat{\got{sl}}_n}                      

\def\hslt{\widehat{\got{sl}}_2}

\def\1{1\!\!1}

\def\tp{\otimes}

\def\Li{\Lambda_i}
\def\bLi{\bar{\Lambda_i}}

\def\bL{\bar{\Lambda}}
\def\GL{{\rm GL}_{\infty}}
\def\vac#1{|#1>}
\begin{document}

\thispagestyle{empty}
\setcounter{page}0
 \begin{titlepage}
\bigskip\bigskip
\begin{center}
{\Large\bf Intertwining Operators and Soliton Equations}\\
\bigskip
\bigskip
{\large M. Golenishcheva-Kutuzova \footnote{E-mail:
kutuzova@vitep5.itep.ru ;  kutuzova@desargues.univ-lyon1.fr}} \\
\bigskip
{\it International Center for Nonlinear Science of Landau Instutute
of Theoretical Physics,\\ 117 940, Moscow, Russia}\\
\bigskip
{\large D. Lebedev \footnote{E-mail: lebedev@vitep5.itep.ru ;
lebedev@desargues.univ-lyon1.fr}}\\
\bigskip
{\it Institute of Theoretical \& Experimental Physics,
117259 Moscow, Russia}
\bigskip
\end{center}
\bigskip

\bigskip
\bigskip

\begin{abstract}
 In this paper we generalize the fermionic approach to
the KP hierarchy suggested in the papers of Kyoto school 1981-1984 (Sato,
Date, Jimbo, Kashiwara and Miwa).
 The main idea is that
the components of the intertwining operators are in some sense a
generalization of free fermions for $gl_{\infty}$. We formulate in terms of
intertwining operators the integrable hierarchies related to Kac-Moody Lie
algebra symmetries.  We write down explicitly the bosonization of these
operators for different choices of Heisenberg subalgebras. These different
realizations lead to different hierarchies of soliton equations.  For
example, for $sl_N$-symmetries we get  hierarchies obtained as $(n_1,...,
n_s)$-reduction from $s$-component KP $(n_1+...+n_s = N)$ introduced by V.Kac
and J.van de Leur.
 \end{abstract}
\end{titlepage}

\newpage
\tableofcontents

\section{Introduction}

It is well known that there is a deep connection between the theory
of soliton equations and  classical infinite dimension Lie algebras.  This
relation was discovered first in the paper of M.Sato \cite{S} and than was
developed in the paper of Kyoto school \cite{DJKM1}-\cite{DJKM2},
 using the boson-fermion
correspondence in 2-dim QFT.
The main idea can be illustrated in the example of $KP$ equation.
One can interpret the set of solutions of
the KP hierarchy as an
orbit $\got{O} =\GL |0>$ of the group $ \GL$ of the vacuum vector
$|0>$  in the basic representation $V(\Lambda_{0})$ of $\GL$ .
The system of equations defining $\got{O}$ can be determined in terms of
operator
$$
S =  Res|_{z=0} \Psi(z)   \tp \Psi^* (z),
$$
where $\Psi(z)=\sum_{i\in \ZZ} \Psi_i \, z^{-i-\frac12}, \; \;
  \Psi^*(z)=\sum_{i\in \ZZ}\Psi_i^* \, z^{-i-\frac12}$ are charged fermionic
fields. The operator $S$ commutes with the diagonal action of $\GL$ on
$\got F \tp \got F$, where$ \got F$ is a fermionic Fock space, and
 $S (|0> \tp |0>  ) = 0 $. It follows that any element
$\tau \in  \got O$
 satisfies the equation
 \be \label{k1}
S(\tau \tp \tau )=0.
\ee
Using the boson-fermion correspondence, one can rewrite this equation as a
system of partial differential equations in the Hirota bilinear form.
Thus, we get the KP hierarchy. Using the so-called reduction procedure,
 one can write down a hierarchy of Hirota bilinear equations
for the orbit of the highest weight vector in the basic representation of
the loop group $SL_n$. This gives  the KdV hierarchy for $n=2$,
the Boussinesq hierarchy for $n=3$, e.t.c. The $[n_1,n_2, ...,n_s]$ reduction
from the s-component KP hierarchy \cite{KV} - \cite{V} leads to
the hierarchies
corresponding to the different realization $R_w$  of the basic
representation of the affine Kac-Moody Lie algebra $\hsln, \; \, n =
n_1+...+n_s$.  It is known that all these realizations are parametrized by
conjugacy classes of the Weyl group $W$ of $\hg$ \cite{KP}. A similar
approach can be applied to the group $O_{\infty}$ \cite{KV}- \cite{V}.

The problem how to construct the hierarchies associated to arbitrary loop
groups in a unified fashion, including the exceptional ones, was developed by
V.Kac and M.Wakimoto in \cite{KW}, using the quadratic Casimir operator
in place of the fermionic-like Casimir operator
$S$. In fact, the equation (\ref{k1})
is equivalent to $S^*S(\tau \tp \tau)=0$, so one may consider the operator
$\Omega=S^*\,S $ instead of $S$, which is the Casimir operator for $\GL$. This
operator allows the generalization for the general affine Lie algebra $\g$ in
the following way. Let $< \cdot \,|\, \cdot>$ be an invariant symmetric
nondegenerate bilinear form on $\hg$, and let $V$ be a representation of
$\hg$ that "integrates" to a representation of the corresponding loop group
$\hat G$. Let $\{u_i\}$ and $\{u^i\}$ are  the dual basis of $\hg$, i.e.
$< u_i|\, u^j> = \delta_{ij}$. Then the operator
$\Omega=\sum\limits_j u_j \tp u^j $
commutes with the diagonal action of $\hat G$ on $V \tp V$. Let $v_0 \in V$
be such that $\Omega(v_0 \tp v_0)=\lambda(v_0 \tp v_0)$, $\lambda \in \CC$.
Then any $\tau$ from the orbit $G \cdot v_0$ satisfies the equation
\be\label{k2}
\Omega(\tau \tp \tau)=\lambda(\tau \tp \tau)
\ee

The main idea of this paper came from  \cite{GKLMM}, where
the analogue  of the fermionic-like Casimir
operator was constructed in terms of the intertwining operators. For the
level one modules, these operators form a closed operator algebra  of the
parafermionic type  similar to \cite{FZ} and from some point
of view can be interpretated
as the generalization of  fermion operators. Namely, for $\GL$ case,
the fermionic
fields $\Psi(z)$ and $\Psi^*(z)$  can be interpreted as the intertwining
operators for  $\GL$ level one (fundamental) modules.

The classical hierarchies, related with the representations of Kac-Moody Lie
algebras expressed in terms of the intertwining operators, also allow the
q-analogues for quantum intertwining operators for the corresponding quantum
group. The first steps in this direction were done in  \cite{GKLMM}.
This paper present the general approach how to build up the
analogue of the operator $S$ in classical and quantum cases. For quantum
$U_q(sl_2)$ the corresponding hierarchies of bilinear equations
were written.
These equations are combinations of differential and q-difference equations.
 The attempts to extend this approach to the general case of the quantized
unveloping algebra meet with obstacles. Roughly speaking, these obstacles are
related to the following  problem:
 what is the
quantum evolution on quantum Grassmannian. Some investigations in this
direction were done  in the papers  \cite{GKLMM} -\cite{KKL},
 but it does not
allow to rewrite the abstract equation (\ref{k1}) on the $U_q\g$ -orbit of the
vacuum vector in the Hirota-like appropriate form. Anyway, there is no
obstruction to use the intertwining operator approach in the classical case.
For affine Kac-Moody Lie algebras this approach
gives some addition possibilities to compare with the
 quadratic - Casimir one. The exceptions are provided
by the cases when the dimension of the space of such operators is 0, but in
general, in this approach one can consider the equations related with the
tensor products of different $\hg$-modules and not necessarily of level one.
In this paper we explain how to write down  Hirota bilinear
equations on the $\tau$-function of the type
\be
S(\tau \tp \tau) = 0,
\ee
where $S$ is the  fermionic-like Casimir operator for the affine Lie
algebra $\hg$ expressed in terms of intertwinings operators between
the level one modules. The case of the Wakimoto modules symmetries
will be considered
in \cite{GK}.
The intertwining operators  are the central object of the representation
theory of affine Lie algebras and associated conformal field theories.
Remind some facts about the intertwining operators \cite{FR}.
To any representation $V$ of $\g$ and a formal variable $z$, one can
associate two types of representations. One type is the highest weight
representation $V_{\lambda, k}$ of level k (equal to the value of the
central element c). This representation has $\ZZ$-graded structure
$V_{\lambda, k} = \bigoplus_{n \in \ZZ}   V_{\lambda, k}[-n]$ compatible
with the one of $\g$ and its top subspace $V_{\lambda, k}[0]$ is isomorphic
 to  $V_{\lambda}$. The second type is a finite dimensional evaluation
representation for which $k = 0$.
  As a vector space $V[z] \cong V \tp
\CC((z))$, where $\CC((z))$ stands for the Laurent series in $z$, and the
generators $J^n_x = x \tp t^n, \; \; x\in \g$ of $\hg$ acts by $x \tp z^n$
on $V[z]$.
Introduce the fields in $\hg$;
$$
J_x(z) = \sum_{n \in \ZZ} J^{n}_x \, z^{-n -1}
$$
The commutation relations admit the following form
\be\label{cur}
[ J_x(z) , \: J_y(w) ] =  J_{[x , y]}(w) \,\delta (z - w) +
\delta^{'}(z - w) c,
\ee
where $  \delta (z - w) = z^{-1} \sum_{n \in \ZZ}\left(
\frac{z}{w}\right)^n$.
Consider the intertwining operators
\be\label{tw}
        \Psi(z) :  V_{\mu, k} \longrightarrow
V_{\nu, k } \tp V_{\lambda}[z]
\ee
where $\mu, \nu, \lambda \in P_+$. As usual in (\ref{tw}) one makes
some shift in grading by $z$, coming from the grading of representation
see (\cite{FR}).
A fixed choice of an element $v \in
V_{\lambda}^* $ gives rise to the operator
\be\label{t3}
        \Psi_v (z):  V_{\mu, k} \longrightarrow
V_{\nu, k } \tp \CC [[z]],
\ee
or in the component form
$$
\Psi_v (z) = \sum_{n \in \ZZ}\Psi_v [n] \, z^{-n}.
$$
Since the intertwining property of $\Psi(z)$ does not depend on
multiplication by any power series of $z$ we will choose the grading
in (\ref{t3}) consistent with the graded structures of the representation.

By definition, the intertwining operator $\Psi_v (z)$ satisfies
\be\label{t5}
[ J_x(z), \: \Psi_v (w)]=  \Psi_{xv} (w) \, \delta (z - w)
\ee
For the adjoint vector representation $V_{\lambda}^* $ one
can define in the same fashion the adjoint intertwining operator:
\be\label{ad}
        \Psi^*(z) :  V_{\mu', k} \longrightarrow
 V_{\lambda}^*[z] \tp V_{\nu', k } ,
\ee
and for each vector $v \in V_{\lambda}$, the components of the
adjoint operator:
\be\label{ad1}
 \Psi(z)_v^* :  V_{\mu', k} \longrightarrow
V_{\nu', k } \tp \CC [[z]].
 \ee
 Consider the dual basises $\{v_i\}$ and $\{v^i\}$
in $V_{\lambda} $  and  $V_{\lambda}^* $  with respect to
a natural pairing  $V_{\lambda} \tp V_{\lambda}^*  \to \CC$.
Then the operator
\be\label{ad3}
K (z) = \sum_{i} \Psi_{v^i} (z) \tp \Psi_{v_i}^* (z) :
V_{\mu, k} \tp V_{\mu', k} \longmapsto
\ee
$$
\longmapsto V_{\nu, k } \tp  V_{\nu',\, k } \tp \CC((z))
$$
commutes with the diagonal action of $\hg$ on the tensor product:
$$
\Delta \hg \: K(z) = K(z) \:
 \Delta \hg
$$
and doesn't depend on the choice of the dual basis on
 $V_{\lambda} $  and  $V_{\lambda}^* $.
This operator is a fermionic analogues of the operator (\ref{k1}).

 \begin{de}
 Consider the following data:
\end{de}
  (i)  affine
Kac-Moody Lie algebra $\hg$\\
(ii) Casimir operator $K(z)$ (\ref{ad3})\\
(iii)  vertex operator construction $R$ (bosonization) of
the operators $\Psi_{v^i} (w) $ and $\Psi_{v_i}^* (w)$ and the
fields $J_x(z)$ of $\g$.\\
To these data we will associate the hierarchies of soliton equation,
which we will call the hierarchy associated with the data (i)-(iii).

Note that for given modules our construction depends on the dimension
of the space of the intertwining operators and on the vertex operator
construction $R$. It is known, that for simply-laced case  and the
level one representations such  constructions are
parametrized by the conjugacy classes of the Weyl group $W$ of $\hg$.

We would like to mention the series of papers
\cite{W,GHM,BGHM,A} where different aspects
of application of the Drinfeld-Sokolov
procedure in the case of intermediate gradation of Affine
Lie algebras were constructed. It will be useful to find
connection between this approach and ours (we thank H.Aratyn for drawing
our attention to the papers \cite{W,GHM,BGHM,A}).

\section{Intertwining Operators of level one modules.\\
Homogeneous construction }
 In this section we give the bosonization of intertwining operators for level
 one modules $V(\Li),\,i\in~I$. All these operators can be realized in
the  space
$$ V_P=\bigoplus \limits_{i\in I} V(\Li),
$$
the sum is taken over the all level one fundamental modules of $\hg$. These
operators form an operator algebra of "parafermiomic type". To give
explicitly this construction we  introduce the notion of vertex
operator algebra associated to the weight lattice $P$. This vertex algebra
can be considered as a generalization of the n-fermionic Fock space for $\gli$.

\subsection{Vertex algebra $V_P$ associated with the weight lattice $P$}
Let $\g$ be a simple finite-dimensional Lie algebra of type
$A_n,\,D_n,\,E_6,\,E_7$(case $E_8$ is exceptional for our construction for the
reason that the root lattice of $E_8$ is self-dual). Let $Q$ be the
root lattice of $\g$ with the
ordered basis $\{\alpha_i\},\,i=1,...,l$. We fix the normalization of
 the symmetric bilinear form $(\cdot|\cdot)$  on $\g$ by
$(\alpha_i|\alpha_i)=2$. Let $\hg=\g\tp\CC[t,t^{-1}]|\oplus \CC c$
be the
corresponding affine Lie algebra associated to $\g$ and let
 $\Lambda_i,\,i=0,1,...,l$
be the fundamental weight of $\hg$. We have (see  for
notations \cite{K1}) $\Li=\bLi + a_i^{\vee}\Lambda_0$, where
$\bL_0=0$ and $\bL_1,...,\bL_l$ are the fundamental weights of
$\g$ and $a_0^{\vee}, ... , a_l^{\vee} $
are the Coxeter numbers of $\hg$.
 The level of the representation with the hiest weight
$\Li$ is one if and only if
$a_i^{\vee}=1$.
Let denote by $I$ the set of level one fundamental modules:
$I=\{i\in \{0,...,l\}\, | a_i^{\vee}=1\}$.
Let $P$ be the weight lattice
of $\g$. It can be proved that $P$ can be represented as
 \be\label{P}
 P= \bigsqcup_{i\in I} \:\{\bL_i + Q\}.
\ee
We can extend the
bilinear form $(\cdot|\cdot)$  from $Q$ to $P$ .

 The twisted group algebra
$\CC_{\ve}[P]$
 is an algebra with the basis
$\{e^{\al},\: \al \in P \}$ and  "twisted" multiplication:
\be\label{coc}
e^{\al} \cdot e^{\beta}\, = \, \ve (\al, \beta) e^{\al + \beta} ,
\ee
where $\ve$ is a 2-cocycle on $P$ defined by

\begin{eqnarray}
&&1. \: \ve (\al, 0)=\ve(0, \al)=1
\label{coc1a}\\
&&2. \: \ve (\al,\beta)= (-1)^{(\al, \beta)} \ve (\beta, \al), \: \:
{\rm for}\, \al
\in Q, \: \beta \in P
\label{coc1b} \\
&&3. \: \ve (\beta , \gamma)\ve (\beta + \gamma , \al
)=\ve (\beta , \gamma + \al ) \ve (\gamma , \al ).
\label{coc1c}
\end{eqnarray}

Given a 2-cocycle
$\ve : P \times P \to \CC^*$ , one associates to $\ve$ a function $B_{\ve} :
P \times P \to \CC^*$ defined by
\be\label{coc2}
B_{\ve}(\al , \beta ) = \ve
( \al , \beta) \ve (\beta, \al )^{-1}
\ee
It is clear that $B_{\ve}$ is
skewsymmetric, i.e.,
\be\label{coc2a}
B_{\ve}(\al , \beta ) = B_{\ve}^{-1}
  (\beta , \al ) .
\ee
And (\ref{coc}) is equivalent to
\be\label{coc4}

e^{\al} e^{\beta} = B_{\ve}(\al , \beta )e^{\beta} e^{\al}
\ee
Using the associativity, we see that $B_{\ve}$ is bimultiplicative, i.e.,
\be\label{coc3}
 B_{\ve}(\al + \gamma , \beta ) = B_{\ve}(\al , \beta )  B_{\ve}(\gamma ,
\beta ), \:{\rm and} \: \:  B_{\ve}(\beta , \al +\gamma ) = B_{\ve}(\beta ,
\al ) B_{\ve}( \beta , \gamma ).
\ee
It is clear that bimultiplicativity implies the cocycle property
(\ref{coc1a}).  For the weight lattice $P$ under consideration we would like
to construct explicitly a bimultiplicative cocycle $\ve (\al,\beta)$ on $P
\times P$  with the properties (\ref{coc1a})- (\ref{coc1c}).  Unfortunately,
in the general case it is impossible. Let $N$ be the index of $Q$ in $P$. For
 $N$ odd such a cocycle exists, for $N$ even it is possible to construct a
 such cocycle only on $Q \times P$. Note that the property (\ref{coc1b}) is
 responsible for a right commutator's relations between the fields of
 Kac-Moody Lie algebra $\hg$ and between  the field of $\hg$ and the
 components of the intertwining operators.  First, construct explicitly the
 cocycle $\ve ( \al , \beta)$ on the integral lattice $Q \in Q$ with  values
$\pm 1$. Choose an ordered basis $\al_1,..., \al_{\ell}$ of $Q$ over $\ZZ$ ,
 let

\bse\label{coc6}
\be\label{coc6a}
 \ve ( \al_i , \al_j ) = (-1)^{(\al_i | \al_j)}, \ \ \ \ \ {\rm if} \, i<j;
\ee
\be\label{coc6b}
\ve (\al_i , \al_j ) = 1, \ \ \ \ \   {\rm if} \;  i>j
\ee
\be\label{coc6c}
 \ve ( \al_i, \al_j )= (-1)^{(\al_i | \al_i)/2}, \ \ \ \ \  {\rm if} \; i=j
\ee
\ese
 and extend  to $Q \times Q$ by bimultiplicativity. We obtain a
 bimultiplicative function $\ve : Q \times Q \to \{\pm 1\}$. The
 bimultiplicativity implies the cocycle properties (\ref{coc1c}).
 Preserving the multiplicativity we would like to
   extend this cocycle to the lattice. Introduce the following matrices
 \be\label{coc9}
 B=\left(b_{i,j} \right),   \; \; b_{i,j}=\frac{1}{i\pi}\log
\ve ( \al_i, \al_j ) ,\\
 C=\left(c_{i,j} \right) , \; \; c_{i,j}=\frac{1}{i\pi}\log
\ve ( \al_i, \bLi_j ) ,\\
C^{'}=\left(c^{'}_{i,j} \right) , \; \; c^{'}_{i,j}=\frac{1}{i\pi}\log
\ve ( \bLi, \al_j ) ,\\
D=\left(d_{i,j} \right) , \; \; d_{i,j}=\frac{1}{i\pi}\log
\ve ( \bLi, \bL_j ).
\ee
The relation (\ref{coc1b}) is equivalent to
$$
B=B^t + A +2L, \; \; \;C^{'} - C^t = E + 2K, \; \;\\
B=CA, \; \; \; B^{'} =AC , \; \; \; B =AD+ 2S, \; \; B^{'} =DA + 2R
$$
and from the bimultiplicativity we have
$$
B =AD+ 2S, \; \; B^{'} = DA + 2R,
$$
where $A$ is the Cartan matrix of $\hg$ , $L$ , $K$,$S$ and $R$ are the matrice
with the matrix elements in $\ZZ$.
From (\ref{coc6a}) - (\ref{coc6c}) $B$ is uniquely defined up to
an even matrix. Choose
$B$ satisfying $B -B^t = A - 2B^t$. Define  $C$ and $C^t$ by
$$
C^{'}=A^{-1}(B + 2B^t), \; \; C =BA^{-1}.
$$
This give us a well defined cocycle on $Q \times P$ or $P \times Q$
with the property (\ref{coc1b}).

Remark that for even $N=\det A$, $NA$ and $A$ are equal up
to the even
matrix.  Than, $NB$ satisfies the same relation as $B$ and  $NB^t
A^{-1}$ is a  matrix with the integer elements. Define $D$ by
$$
 D = A^{-1}(B + 2B^t)A^{-1}
$$
 Then all relations are satisfied and we get by
bimultiplicativity the cocycle on $P \tp P $ with the property
 (\ref{coc1a})-(\ref{coc1c}). For $N$ odd, it is impossible
 to find the solution
for  the matrix $D$ (we need that $B^tA^{-1}$ be an integer matrix).
 Such cocycle does not exist on $P\tp P$ for $\hslt$. This
case will be considered in the (\ref{sl2}).

Let $\h = \CC \tp_{\ZZ}Q$
 be the complexification of $Q$, and extend the bilinear form $(\cdot |
\cdot )$ from $Q$ to $\h$ by bilinearity. Let
\be\label{fu}
\hat{\h} = \h \tp
\CC [t, t^{-1}] + \CC k \ee be the affinization of $\h $ (Heisenberg
subalgebra). Let $S$ be the symmetric algebra over the space \be \h^{<0} =
\sum_{j<0} \h \tp t^j.
\ee
We shall write $ht^j$ in place of $h \tp t^j$ for
short.

We define the space of states of the vertex algebra that we shall associate
to the lattice $P$ as
\be\label{s1}
V_P = S \tp \CC_{\ve} [P],
\ee
where $S$ is the symmetric algebra of $h^{<0}$. The vacuum vector is
$\vac{0} = 1 \tp 1$. The action $\pi$ of $\hat{\h}$ on $V_P$ is defined as
follows. Let $\pi = \pi_1 \tp 1 + 1 \tp\pi_2$, where $\pi_1$ is the action on
$S$, and $\pi_2$ is the action on $\CC_{\ve} [P]$ defined by letting
$\pi_1(c)=I, \:  \pi_1(ht^n)$ be the operator of multiplication by  $ht^n$ if
$n<0, \: \pi(ht^n)$ be the derivation of the algebra $S$ defined by
$(h_1t^n)(h_2t^{-s}) = n \delta_{n,s}(h_1|h_2)$ if $n>0$ and $\pi_1(h)=0$.
$\pi_1(ht^n)$ acts trivially on $\CC_{\ve} [P]$ if $n \ne 0$, and
$\pi_2(h)e^{\al} = (h|\al)e^{\al}, \: \pi_2(c) = 0$. Denote by $h_n =
\pi(ht^n), \: h\in \h, n \in \ZZ$ and consider the following {\,} End
$V_P$-valued fields

\be\label{z1}
h(z) = \sum_{n\in\ZZ} h_n z^{-n-1},
\ee
and the correspoding Veneziano field $\tilde h (z)$
\be\label{z2}
\tilde h (z) =-ih - i h_0 \log z + i\sum_{n\ne 0}\frac{z^{-n}}{n} h_n ,
\ee
so that
$h(z) = i\partial \tilde h (z)$.
 We have
$[h_m, h_n^{'}] = m \delta _{m, -n} (h|h^{'}), \: h,h^{'}\in \h, \:m,n
\in\ZZ$.  Denote by $e^{\al}$ the operator on $V_P$ of multiplication by
$1\tp e^{\al}, \: \al \in P$,
we have
$[h_n, e^{\al}] =\delta _{n, 0}(\al | h) e^{\al}$.
Let $Y(v, z), \: z \in \CC$ be the field, corresponding to
the vector $v \in V_P$ via the state-field correspondence (for more detailes
about Vertex Operators Algebra (VOA) see \cite {K1}). In particularly, this
means that $Y(v, z) \vac{0}|_{z=0} = v$. Let
$$
\Gamma_{\al}(z) = Y(1 \tp e^{\al}, z) \: \: \; {\rm for} \:\; \al \in P .
$$
From the commutators relations with $h_n$ we derive that
\be\label{V1}
 \: \Gamma_{\al}(z) =
e^{\sum_{n>0}
 \frac{z^n}{n}
 \al_{-n}}e^{- \sum_{n>0}
\frac{z^{-n}}{n}
\al_{n}}
\tp e^{\al}z^{\al_0},
\ee
 or in a short form
$
\Gamma_{\al}(z) = e^{i \tilde h (z) }.
$
We have the following OPE's
\be\label{V2}
 \: h(z)\Gamma_{\al}(w) \sim \frac{(\al|h)\Gamma_{\al}(w)}{z-w}, \:\:
h\in \h, \: \al \in P,
\ee
\be\label{V3}
 \Gamma_{\al}(z)\Gamma_{\beta}(w) = \ve (\al, \beta) (z - w)^{(\al| \beta)}
 \Gamma_{\al,\beta}(z, w), \:\: \al,\beta \in P, \: |z|>|w|,
\ee
where
$$
\:\: \Gamma_{\al,\beta}(z, w) =  e^{\sum_{n>0} \frac{z^n}{n} \al_{-n}+
\frac{w^n}{n} \beta_{-n}}
  e^{-\sum_{n>0} \frac{z^{-n}}{n} \al_{n}+
\frac{w^{-n}}{n} \beta_n} \tp e^{\al+\beta}z^{\al_0}w^{\beta_0}.
$$
From (\ref{V3}) and the cocycle properties (\ref{coc1b}) we get
that for $\al \in Q, \: \beta \in P$
the fields $\Gamma_{\al}(z)$ and $\Gamma_{\beta}(w)$ are pairwise local in
 the strong sense \cite{K1} and for the general case satisfy the OPE of the
parafermionic type \cite{FZ}. Due to (\ref{P}),
\be\label{P2}
V_P = \bigoplus_{i \in I} V(\Lambda_i).
\ee
as a linear space.
In particular, $V(\Lambda_0) \simeq V_Q = S \tp \CC_{\ve}[Q]$ is a rational
VOA and $V_P$ is a direct sum of the VOA $V(\Lambda_0)$ and all
its representations (see \cite{K1}) .

Let $X_{\al}, \: h_{\al}, \: \al \in \Delta$ be the Cartan basis of $\g$.
Define the fields
$$
X_{\al}(z) = \sum_{n \in \ZZ} X_{\al}t^n \, z^{-n-1},\:\:\: \;
h_{\al}(z) = \sum_{n \in \ZZ} h_{\al}t^n \,  z^{-n-1}.
$$
Then the correspondence
\be\label{V4}
 X_{\al}(z)\longmapsto \Gamma_{\al}(z), \:\:  h_{\al}(z)\longmapsto \al (z),
\:\:  c \longmapsto 1
\ee
 gives us the vertex operator representation of $\hg$
(Frenkel-Kac construction \cite{KKLW}) Operators (\ref{V4}) preserve the
decomposition (\ref{P2}), namely in $V(\Lambda_0)$ we have the basic
representation in homogeneous realization.
\subsection{Bosonization of level one intertwining operators}

Let again $V(\Li) , \: i \in I$ be the level one fundamental weights of $\hg$,
and let $V(\bLi)[z]$ be the corresponding
evaluation module for $\hg$. For the  dual module $V^*(\bLi)[z]$  we have
$V^*(\bLi)[z] \simeq V^*(\bL_j)[z]$ for some $j \in I$. Denote by $P(\bLi )$
the set of weights of the finite dimensional $\g$ -module $V(\bLi)$ and let
$\{e_{\gamma}\}, \:  \gamma \in P(\bLi)$ be the weight basis of $V(\bLi)$
normalized as follows
 \be\label{ba1}
  e_{\gamma} =X_{-\al}e_{\bLi}, \:\: {\rm
where}\;\;\; \gamma = \bLi - \al,\;\; \al \in \Delta_+.
 \ee
  Define the dual basis $\{e_{\gamma}^*\} , \;\; \gamma \in P(\bLi)$ in
 $V^*(\bLi)$  by
\be\label{ba2}
e_{\gamma}^* =(-1)^{(\bLi|\bLi-\gamma )}
\ve (\al , \al)\,X _{\al}e_{-\bLi} \in V^*(\bLi).
\ee
Consider the intertwining
operator
 \be\label{i1}
\Phi^i(z) :  V(\Lambda_m)\longrightarrow
V(\Lambda_n)\tp V^*(\bLi)[z] z^{(\bL_m|\bL_i)} \ee and the adjoint
 intertwining operator
 \be\label{i2}
\Psi^i(z) :  V(\Lambda_p)\longrightarrow
V(\Lambda_s)\tp V(\bLi)[z] \,z^{-(\bL_p|\bL_i)},
\ee
where $\bL_m +\bLi \in
\{\bL_n + Q\}$ and $\bL_p - \bLi \in \{\bL_s + Q\}$.  Each vector $e \in
V(\bLi) $ defines the component of $\Phi^i(z)$
\be\label{i3}
\Phi_e^i(z) :
 V(\Lambda_m)\longrightarrow V(\Lambda_n) \tp \CC ((z))z^{(\bL_m|\bL_i)},
\ee
and $e^* \in
V^*(\bLi)$, the components of $\Phi_e^i(z)$
\be\label{i4}
\Psi_{e^*}^i(z) :
 V(\Lambda_p)\longrightarrow V(\Lambda_s) \tp \CC ((z))z^{-(\bL_p|\bL_i)}.
 \ee
For the weight
basis $\{e_{\gamma}\}$ and $  \{e_{\gamma}^*\}, \: \gamma \in P(V(\bLi)),$ on
$V(\bLi) \:$ and $ V^*(\bLi))$ respectively  we will write
$$
\Phi_{\gamma}^i(z) =
\Phi_{e_{\gamma}}^i (z), \; \; \:\: \Psi_{\gamma}^i(z) =\Phi_{e_{\gamma}^*}^i
 (z)
$$
 From (\ref{V1}) -(\ref{V2}) and the normalization given by
\be\label{n1}
\Phi_{e_{\bLi}^i} (z) \vac{0}|_{z=0}= 1 \tp e^{\bLi} , \:\:
\Psi_{e_{\bLi^*}^i} (z) \vac{0}|_{z=0}= 1 \tp e^{-\bLi}
\ee
 we get
\be\label{b1}
\Phi_{\gamma}^i(z) = \ve (-\al , \bLi) \Gamma_{\gamma}(z) \:\:   \:\\
\Psi_{\gamma}^i(z) = (-1)^{(\bLi, \al)}\ve(\al, -\bLi) \ve (\al ,\al)
 \Gamma_{- \gamma}(z),
\ee
where $\gamma \in P(\bLi)$ and $ \gamma = \bLi - \al $. Indeed, for
$\gamma\in P(\bLi)$
   we have $\gamma = \bLi - \al , \: \al \in \Delta_+$ and
$(\bLi , \al) = 1$ . From this and  (\ref{V2}) and (\ref{coc6}) we
have the following OPE
 \be\label{b2}
X_{-\al}(z) \Gamma_{\bLi}(z)  \sim \frac
{\ve(- \al , \bLi) \Gamma_{\gamma}(z)}
{(z -w)^{(\al , \bLi)}} , \; \; \;     {\rm for}\; \; |z|>|w|.
\ee

 It is equivalent to the following commutation relations for the intertwining
operators
 \be\label{b3}
[X_\al (z), \Phi_{\bLi}^i(w) ] = \Phi_{\gamma}^i(z) \delta (z - w)  \: \:
{\rm and}\\

[X_{\al} (z), \Psi_{\bLi}^i(w) ]=
 \ve (-\al, \al) \Psi_{\gamma}^i(z) \delta (z - w) ,
\ee
 where  $\Phi_{\gamma}^i(z)$ and
$\Psi_{\gamma}^i(z)$ are given by (\ref{b1}). Due to the cocycle properties
(\ref{coc1b}) and relations (\ref{b3}) we get the more general commutation
relations (\ref{i1}) between the fields $X_\al (z), \: \: \al \in Q$ and
the components $ \Phi_{\gamma}^i(z)$ and  $ \Psi_{\gamma}^i(z) $
of the intertwining operators (\ref{i3}), (\ref{i4})
\be\label{i6}
[X_\beta (z), \Phi_{\gamma}^i(w) ] =
\ve (\beta , -\al )\Phi_{\gamma +\beta}^i(z) \delta (z-w)
  \ee
 $$
 [X_{\beta} (z), \Psi_{\gamma}^i(w) ]=
\ve (\beta , \al )\Psi_{\gamma + \beta}^i(z)\delta(z-w),
$$
which coincide with the defining relations (\ref{t5}).
 \begin{ex}
Consider the case $\hg = \hsl (N, \CC)$, $N >2$.The case $N =2$ will be
considered in the next section.
\end{ex}
In this case all fundamental  modules $V(\Li)$,
$i \in {\rm I} = \{0,\ldots , N-1\}$ are at level one. We have $P=
\bigsqcup_{i\in {\rm I}} \:\{\bLi + Q\} $ , $V^*(\bLi) \simeq
V(\bL_{N-i})$ and the decomposition
$$V_P = \oplus_{i=0}^{N-1} V(\Li).$$
All $V(\Li)$ are invariant under the action of $\hsl (N,
\CC)$ given by
\be\label{sl1}
X_{\al} \longmapsto \Gamma_{\al}(z); \: h_{\al} \longmapsto \al (z), \:
\al \in \Delta; \:c \longmapsto 1
\ee
and
$$
 \Phi_{\gamma}^i(z)= \ve (- \bLi +\gamma , \bLi) \Gamma_{\gamma}(z), \:\;
\Psi_{\gamma}^i(z)= (-1)^{ (\bLi - \gamma, \bLi)} \ve( \bLi - \gamma, \gamma)
\Gamma_{-\gamma}(z)
$$
 for $\gamma \in P(\bLi)$.
Note that from the definition of the cocycle,
$\Psi_{\gamma}^i(z)$ are well defined operators on $V_P$ only for $N$ odd
and for $N$ even only on $V(\Lambda_0)\simeq V_Q$.
\section{Hirota bilinear equation in terms of intertwining\\ operators.
Homogeneous construction}

Let $ \Phi_{\gamma}^i(z)$  and
$\Psi_{\gamma}^i(z)$ be the components of the intertwining operators
$\Phi(z)$ and   $ \Psi(z)$ , defined via the weight basis
$\{e_{\gamma}\}$ and
$\{e^*_{\gamma}\}$ see (\ref{b1})-(\ref{b2}) in $V(\bLi)$ and $V^*(\bLi)$
respectively.

Consider the operator
\be\label{u1}
K^i(z) =\sum_{\gamma \in P(\bLi)} \Phi_{\gamma}^i(z) \tp  \Psi_{\gamma}^i(z):
V(\Lambda_m) \tp V(\Lambda_p) \longrightarrow V(\Lambda_n) \tp
V(\Lambda_s) \CC[z,z^{-1}]\,z^{(\bLi , \bL_m - \bL_p)},
\ee
where $m,p,n,s \in I$ and the highest weights are related as in (\ref{i1}) -
(\ref{i2}).
 By the definition 1.1  of the introduction, to construct the
integrable hierarchy of soliton equations we need the following data.\\
a)  operator $K : V(\Lambda_m) \tp V(\Lambda_p) \longrightarrow
V(\Lambda_n) \tp V(\Lambda_s) $,
commuting with the diagonal action of Lie algebra $\hg$ (corresponding Lie
group $\hat G$):
$  \Delta \hg K = K \Delta \hg$,
where $\Delta $ is a natural action of the $ \hg\; \; \;(\hat G)$ on
the tensor product of two modules.\\
b)   vertex operator construction of
 $\hg$ and of $K$.\\
c)  condition $K (\vac{v_m} \tp \vac{v_p}) = 0)$ for the
highest weight vector $\vac{v_m} \tp \,\vac{v_p} \in V(\Lambda_m) \tp
V(\Lambda_p) $.\\
Then the equation
\be\label{u2}
K (\tau^{(m)} \tp \tau^{(p)}) = 0, \:\:\:\; \tau^{(m)} \tp \tau^{(p)}  \in
V(\Lambda_m) \tp V(\Lambda_p)
\ee
determined the orbit of the vector $\tau^{(m)} \tp \tau^{(p)}$ under the action
of $\hat G$. If the operator $K$ exists and can be presented in  bosonic form,
we can
 rewrite this equation as a hierarchy of soliton equation in Hirota
bilinear form \cite{DJKM1}, \cite{KW}.

On  $V(\Lambda_m) \tp V(\Lambda_p)$ operator $K^i(z)$ has a decomposition
\be\label{u3}
K^i(z) = \sum_{k \in \ZZ} K_k^i\, z^{-k + (\bLi|\bL_m - \bL_p)}.
\ee
Define the m-vacuum
$\vac{m} = 1 \tp e^{\bL_m}$
for the highest vector on
$V(\Lambda_m)$. For
$$
k> \max_{\gamma \in P(\bLi)}(\bLi - \gamma| \bL_m -
\bL_p)
$$
we have
$$
K_k^i \,(\vac{m} \tp \vac{p}) = 0
$$
and equation
\be\label{u4}
K_k^i \,(\tau^{(m)} \tp \tau^{(p)}) = 0, \:\; \; \tau^{(m)} \tp \tau^{(p)}
\in V(\Lambda_m) \tp V(\Lambda_p)
\ee
determines the hierarchy of soliton
equations. If $\tau \in V(\Lambda_m)$, then
 \be\label{t1}
\tau = \sum_{\beta
 \in Q} \tau_{\beta} \tp e^{\bL_m +\beta},
\ee
where $\tau_{\beta} \in S$. We
can identify the space $S$ with the space $\CC [x_n^{\al}], \: \al \in \Pi,
\: n = 1, 2, ... $, where $\Pi$ is a basis in $Q$. The action on $\CC
[x_n^{\al}]$ of $\al_m,\:\; \al \in P, \:\; m ,n \in \ZZ$ is a naturally
defined representation of the Heisenberg algebra
\be\label{rh}
\al_{-m}
\longmapsto m x_m^{\al}, \: \al_m \longmapsto \sum_{\beta \in \Pi} (\al
,\beta) \frac{\partial}{\partial x_m^{\beta}} \: \;\; {\rm for}\: \; m>0
\ee
$$
{\rm  and} \;\;\; \al_0 \longmapsto 0, \: c \longmapsto 1.
 $$
By linearity
 for all $\gamma \in \CC \tp_{\ZZ} Q,\:\ \:  m \in \ZZ \,$ $\gamma_m$  are well
defined operators on $\CC [x_n^{\al}]$. We will use the notation
$$
 \partial_{\gamma_m} = \sum_{\beta \in \Pi}(\gamma | \,\beta)
\frac{\partial}{\partial x_m^{\beta}},  \: \; \; {\rm  for} \; \;
\gamma_m, \: \; m>0.
$$
 Using
the bosonization form (\ref{V1}), (\ref{b1}) we can write (\ref{u4}) in
 Hirota bilinear form.
\be\label{eq1}
Res |_{z=0} z^{k - (\bLi , \bL_m -
\bL_p)-1}\left (\sum_{\gamma \in P(\bLi)} (-1)^{(\bLi , \bLi - \gamma)} \ve
(\bLi - \gamma, -\bLi - \gamma) \Gamma_{\gamma}(z) \tp
\Gamma_{-\gamma}(z)\right) \cdot \\
\left( \sum_{\beta^{'},\delta^{'} \in Q}
\tau_{\beta^{'}}e^{\bL_m + \beta^{'}} \tp \tau_{\delta^{'}}e^{\bL_p +
\delta^{'}}\right) = 0.
\ee
 We can identify $V(\Lambda_m) \simeq  \CC [x_n'^{\al}] \tp \sum_{\beta \in Q}
e^{\bL_m + \beta}$ and $V(\Lambda_p) \simeq  \CC [x_n''^{\al}] \tp \sum_{\beta
\in Q} e^{\bL_p + \beta}$ . Then (\ref{eq1}) can be written as
\be
\label{eq2}
Res |_{z=0} z^{k - (\bLi,\bL_m - \bL_p)-1}\\
\left(\sum_{\gamma \in P(\bLi)}
(-1)^{(\bLi , \bLi - \gamma)} \ve(\bLi - \gamma, -\bLi - \gamma)
\cdot
\ e^{\sum_{n>0} \frac{z^n}{n}(\gamma_{-n}^{'} - \gamma_{-n}^{''})}
\,e^{- \sum_{n>0} \frac{z^{-n}}{n}(\gamma_{n}^{'} - \gamma_{n}^{''})}
\right.\\
\left.\left( \sum_{\beta^{'}, \delta^{'} \in Q}z ^{(\gamma | \bL_m -
\bL_p + \beta^{'} - \delta^{'})}
        \ve(\gamma , \bL_m - \bL_p + \beta^{'} - \delta^{'})
\tau_{\beta^{'}}(x^{'})\tau_{\delta^{'}}(x^{''})e^{\bL_m + \beta^{'} +
\gamma} \tp e^{\bL_p + \delta^{'}- \gamma}\right)\right)= 0.
 \ee
Introduce the new variables
 $$
x_j^{\al} = \frac{1}{2}(x_j'^{\al} + x_j''^{\al}),\; \; \;
y_j^{\al} = \frac{1}{2}(x_j^{'^{\al}} - x_j^{''^{\al}})
$$
so that
 $$
\frac{\partial}{\partial x_j^{\al}} = \frac{\partial} {\partial
x_j^{'{\al}}} + \frac{\partial}{\partial x_j^{''^{\al}}},\;\; \;
\frac{\partial}{\partial y_j^{\al}} = \frac{\partial}
{\partial y_j^{'{\al}}} - \frac{\partial}{\partial y_j^{''{\al}}}.
$$
The operators $2\gamma_{-n} = \gamma_{-n}^{'} - \gamma_{-n}^{''}$ and
$\gamma_{n} = \gamma_{n}^{'} - \gamma_{n}^{''},\; \;  \: \gamma \in P(\bLi),
\:\;  n>0$ are the operators on $\CC [y_n^{\al}], \: \; \al \in \Pi, \: \;
n \in \ZZ$, defined
according  the (\ref{rh}). Then (\ref{eq2}) can be written as
\be\label{eq3}
Res |_{z=0} z^{k-1}\left( \sum_{\gamma \in P(\bLi)}
(-1)^{(\bLi , \bLi - \gamma)}\ve(\bLi - \gamma, -\bLi - \gamma)\cdot
\right.\\
 \left( \sum_{\beta^{'}, \delta^{'} \in Q}
 z^{(\gamma - \bLi,\bL_m - \bL_p) + (\gamma , \beta^{'} - \delta^{'})}
\ve(\gamma , \bL_m + \beta^{'})
 \ve (-\gamma ,\bL_p + \delta^{'}) \cdot \right.\\
\left.\left.\left(\sum_{p,s=0}
^{\infty} p_s(2 \gamma)p_l(-\tilde {\partial}_{\gamma}) z^{s - l} \right)
 \tau_{\beta^{'}}(x + y)\cdot
\tau_{\delta^{'}}(x - y)  e^{\bL_m + \beta^{'} +
\gamma} \tp e^{\bL_p + \delta^{'}- \gamma}\right)\right)= 0,
\ee
where $p_s$ are the Schur pollynomials.
Rewrite this equation in Hirota bilinear form and look at the coefficient
of
$$
e^{\bL_m + \bLi + \beta} \tp e^{\bL_p - \bLi + \delta}, \: \; \; \beta, \delta
\in Q.
$$
We have
\be\label{eq4}
Res |_{z=0} z^{k-1}\left( \sum_{\gamma \in P(\bLi)}
(-1)^{(\bLi | \bLi - \gamma)} \left( \sum_{\beta^{'}, \delta^{'}
 \in Q}
 z^{(\gamma - \bLi|\bL_m - \bL_p)} \, z^{(\gamma | 2\bLi - 2\gamma + \beta
- \delta)} \times \right.\right.\\
\ve(\bLi - \gamma, -\bLi - \gamma) \ve (\gamma
|\bL_m - \bL_p + 2\bLi - 2\gamma + \beta - \delta)
\left(\sum_{p,s = 0}^{\infty} p_s(2
\gamma)p_l(-\tilde {D}_{\gamma}) z^{s - l} \right) \cdot \\
\left.\left.e^{
\sum_{\al \in \Pi,s \in \z_+} y_s^{\al} D_{\al, s} }
\tau_{\bLi + \beta - \gamma} \cdot \tau_{-\bLi + \gamma + \delta}
\right)\right) = 0.
\ee
Here $\tilde {D}_{\gamma}$ stands for $(D_{{\gamma}_1},  \frac{1}{2}
D_{{\gamma}_2}, \frac{1}{3} D_{{\gamma}_3},...)$ and
$D_{{\gamma}_m} =
\partial_{\gamma_m} = \sum_{\al \in \Pi} (\gamma | \al) \frac{\partial}
{\partial x_m^{\al}}$.

Note that all these Hierarchies are invariant with respect to the
translation
$$
\beta \mapsto \beta + \al, \: \delta \mapsto \delta + \al, \:\;\;
\al \in Q.
$$
\begin{rem}
It is not clear whether these hierarchies are equivalent for different
$k> \max_{\gamma \in P(\bLi)}(\bLi - \gamma|\bL_m - \bL_p)$
\end{rem}
\begin{de}
We will call the hierarchies of bilinear equations, given by (\ref{eq4}),
the hierarchies associated with the operator  (\ref{u1}).
\end{de}
\begin{ex}\label{sl2}
Consider for more details the case $\hg = \hslt$
\end{ex}
We have $I=\{0,1\}$, $\; P=Q\bigsqcup \{\bL_{1}+ Q\}$ and
$P(\bL_1) =\{\frac{\al}{2}, -\frac{\al}{2}\}$, were $\al$ is a positive root.
The bosonization of the fields can be given by
\be\label{slt}
X_{\al}(z) =\Gamma_{\al}(z), \,\, h_{\al}(z)=\al(z) ,\\
\Phi_{\pm\frac{\al}{2}} (z)
=\Gamma_{\frac{ \al}{2}}(z)(-1)^{\frac{\al}{2}}\; \; \; \Psi_{\frac{\al}{2}}
(z)= \pm \Gamma_{\pm \frac{\al}{2}}(z)(-1)^{\frac{\al}{2}}.
 \ee
For $m, p\; mod \, 2$ consider the operator (\ref{u1})
commuting with the diagonal action of $\hslt$ on the tensor product of
two fundamental modules
\be\label{sl4}
 K(z)
= (-1)^{\frac{\al}{2}}\Gamma_{\frac{\al}{2}}(z) \tp
(-1)^{\frac{\al}{2}}\Gamma_{-\frac{\al}{2}}(z) -
(-1)^{\frac{\al}{2}}\Gamma_{\frac{\al}{2}}(z) \tp
(-1)^{\frac{\al}{2}}\Gamma_{-\frac{\al}{2}}(z) :\; \\
V(\Lambda_m) \tp V(\Lambda_p) \longrightarrow
V(\Lambda_{m+1}) \tp
V(\Lambda_{p+1})\tp \CC[z,z^{-1}]\,z^{(\bL_1|\bL_m - \bL_p)},
\ee
where $m, p \; mod \: 2$. We have
$$
(\bL_1|\bL_m - \bL_p) =\left\{\begin{array}{l}
0, \; \; {\rm if} \; \; m=p\\
\pm \frac{1}{2}, \; \; {\rm if} \; \; m\ne p
\end{array}
\right.
$$
We identify the space
$V_P= V(\Lambda_0)\oplus V(\Lambda_1)$ with the space
$$
\CC[y_1, y_2, ...] \tp \sum_{r\in \frac{1}{2}\z} \, e^{ r\al}.
$$
A vector $\tau \in V(\Lambda_i)$ can be written as
$\tau =\sum_{n \in \ZZ} \, \tau_n \tp e^{n\al+i\frac{\al}{2}}, \; i=0,1$.
We get on
$V(\Lambda_0) \tp V(\Lambda_1)$
the following system of Hirota bilinear equations
\be\label{h1}
\sum_{j=0}^{\infty} p_j(y)p_{j+s-l +k}(-\tilde D)
 e^{\sum_{s\ge 1}y_sD_s} \tau_s \cdot \tau_{l + \frac{1}{2}} + \\
\;\;\;\; \; \; \; \;              \sum_{j=0}^{\infty}
p_j(-y)p_{j+l-s+k-1}(\tilde D) e^{\sum_{s\ge 1}y_sD_s} \tau_{s+1} \cdot
\tau_{l - \frac{1}{2}} =0
 \ee
 and on $V(\Lambda_0) \tp V(\Lambda_0)$:
\be\label{h2} \sum_{j=0}^{\infty} p_j(y)p_{j+s-l +k}(-\tilde D) e^{\sum_{s\ge
1}y_sD_s} \tau_s \cdot \tau_{l } - \sum_{j=0}^{\infty}
 p_j(-y)p_{j+l-s-k-1}(\tilde D) e^{\sum_{s\ge 1}y_sD_s} \tau_{s+1} \cdot
 \tau_{l - 1} =0,
\ee
where $p_j(x)$ are the Schur polynomial on $(x_1, x_2,
...)$.  Note that this hierarchy is invariant with respect to the translation
$s \to s +r, \; \; l \to l +r, \;\; r \in \ZZ$. Fix, for simplicity, k=1.  We
look at the coefficient of $y^s_1$ and restrict to the case $l-s = 0,\pm 1,
2$. From (\ref{h2}) we get the following simplest non trivial equations
\be\label{h3}
\left\{ \begin{array}{l} D^2_1 \tau_s \cdot \tau_s +
2\tau_{s-1} \cdot \tau_{s+1}=0 \\ (D^2_1 +D_2) \tau_s \cdot \tau_{s+1}=0\\
(D^2_1 + D_2) \tau_{s-1} \cdot \tau_s =0.
\end{array} \right.
\ee
Letting
$x=x_1,\; t=x_2, \; q(x,t) =\frac{\tau_1}{\tau_0}, \;
q^{*}(x,t)=\frac{\tau_{-1}}{\tau_0} , \; u(x,t)= \log \tau_0$.
 Excluding $u$, we get the NLS system:
\be
\left\{
\begin{array}{l}
q_t = q_{xx} -2q^2q^{*},\\
-q_t^{*} = q_{xx}- 2q q^{*^2}.
\end{array}
\right.
\ee
From (\ref{h1}) and (\ref{h2}) and we get
\be\label{h4}
\left\{
\begin{array}{l}
(D^2_1 - D_2)\tau_s \cdot \tau_{s+\frac{1}{2}}=0 \\
(D^2_1 -D_2) \tau_{s+1} \cdot \tau_{s+\frac{1}{2}}=0\\
 (D^2_1 + D_2)\tau_{s} \cdot \tau_{s+1} =0.
\end{array}
\right.
\ee
Letting
$$
q=\frac{\tau_0}{\tau_{\frac{1}{2}}}, \; \;
q^{*}=\frac{\tau_1}{\tau_{\frac{1}{2}}},
$$
we transform equations (\ref{h4})
into the following system on functions $q$ and $q^{*}$:
\be\label{h5}
\left\{
\begin{array}{l}
q_t + q_{xx} -2q_x(\log q^{*})_x=0,\\
-q_t^{*} + q_{xx}^{*}- 2 q^{*}_x(\log q)_x=0.
\end{array}
\right.
\ee
For any sequence of signs, say, $++-\cdot\cdot \cdot $, and numbers
$a_1, a_2, ... \in \CC, \; \; z_1, z_2, ... \in \CC^{*}$ we can
construct a solution of the NLS and of the (\ref{h5}) hierarchies:
\be
(\tau^0)^{++-...}_{a_1,a_2,...;z_1,z_2,...}(x,t) =
...(1+a_3 X_{-\al}(z_3))(1+a_2 X_{\al}(z_2))(1+a_1 X_{\al}(z_1)) 1 \tp 1
\ee
  and
$$
(\tau^1)^{++-...}_{a_1,a_2,...;z_1,z_2,...}(x,t) =
..(1+a_3 X_{-\al}(z_3))(1+a_2 X_{\al}(z_2))(1+a_1 X_{\al}(z_1))1
\tp e^{\frac{\al}{2}}.
$$

\section{Twisted construction of the level one  Intertwining\\ operators.
Related integrables systems. }

 In this section we give the
 bosonization of intertwining operators for the level one twisted
fundamental modules $V^{\omega}(\Li),\,i\in~I$, where $\omega$ is a element
 of the Weyl group $W$ of $\g$, determined up to the conjugacy class.
 Let $Q_w$ be the orthogonal projection of the root lattice $Q$
 on the subspace stable under the action of $w$ in $Q \tp \CC$.
All these operators can be realized in the  space
$$
V_{P_{\omega}}=\bigoplus \limits_{i\in I} V^{\omega}(\Li),
$$
where $P_{\omega}$ is a lattice dual to the lattice $Q_w$.
It is easy to see that $P_{\omega}$ is a orthogonal projection of the weight
lattice $P$  on the subspace stable under the action of $w$ in $Q \tp \CC$.
 The sum is taken over the all level
one fundamental modules of $\hg$ in a twisted realization. These operators
form a closed operator $\Gamma$-conformal algebras \cite{GKK}
of "parafermiomic type". In this section we will write explicitly
the Hirota bilinear equations in terms of "twisted intertwining operators".
\subsection{Description of nonconjugate Heisenberg subalgebras}
It is well known \cite{KW} that different realization of the basic
(fundamentals) representation of $\hg$ leads to  different hierarchies of
soliton equations. The different vertex-operator's constructions depends on
the different choise of the Heisenberg subalgebra in $\hg$, more
appropriately, on the conjugacy class of Heisenberg subalgebres in $\hg$. In
turn, the Heisenberg subalgebras in $\hg$ classified up to  gauge
transformation are parametrized by the conjugacy classes in the Weyl group
$W$ of $\g$ \cite{KP}. We will remind Kac-Peterson
 construcion of the Heisenbeg
subalgebra corresponding to the given element $\omega$ of a Weyl group.
  Realize $\hg \simeq \g [t, t^{-1}] \oplus \CC c$
as a central extension of the map of the circle to the finite dimensional
Lie algebra $\g$. We want to classify all Heisenberg subalgebras $\{\h(t)\}$
in $\hg$ up to the gauge transformation
 \be\label{G}
\h(t) \to g(t) \h(t)g^{-1}(t),
\ee
where $g(t) \in \hat{G}, \; \; t\in S^1$ , $\h(t) : S^1 \to H$  and
$H$ denotes the set of all Cartan subalgebras in $\g$. Fix in $\h(0)$ a root
system, then for $t = 2\pi$, we get the automorphism $w \in W$
from
$$
     \h(t) = g(t) \h(0)g^{-1}(t).
$$
Let $\{\h(t)\}$ be the set of conjugacy classes of Heisenberg subalgebra
in $\hg$. We have
$$
     \{\h(t)\} \longrightarrow W
$$
 This map can be factorized
\be\label{S}
\left.\left.\{\h(t)\}\right/ G^{S^1} \simeq W \right/ Int,
\ee
where $G^{S^1} = \{ g(t)\: |\:  g(t) \in G, \; \; t \in S^1, \; \; g(0) =
g(2\pi) \}$ and $Int$ is the group of  inner automorphisms of $W$. Fix $w \in
W$ and define the map $W \longrightarrow \{\h(t)\} $ :
\be\label{S2}
w \longrightarrow \h_w(t) = g(t)\,  \h(0)\, g^{-1}(t),
\ee
where $g(0) =1, \;
\; g(2\pi) =w$ and $ g(t) \in G, \; \; t \in S^1$.  He have $g(t) \notin
G^{S^1}$ and the map (\ref{G}) can be factorized to the map (\ref{S}).

If $\h'(0)$ and $\h^{''}(0)$ are two Cartan subalgebras in $\g$, then
$\h'(t)_w $ and $\h^{''}(t)_{w} $  are conjugate in $\g$ under the action of
$ G^{S^1}$ and $\h'(t)_w = \h^{''}(t)_{w'}$ for $w$ and $w'$ conjugate
in $W$.
Our choice of subalgebra $\h(0)$ and the  representative
$\left. w \in W\right/Int$ will be motivated by the problem of bosonisation of
the intertwining operators.

Fix $w \in W$ and let m be the order of $w$. Let
$$
\h (0) = \h = \bigoplus_{j \in \ZZ/m\ZZ}  \h _{j}
$$
be the eigenspace decomposition for $w$, where $\h _{k}$ stands for
the eigenspace attached to  $e^{-2\pi ik}$, so that $\h_0$ is a fixed
point set of $w$. Let $\sigma$ be a lifting of $w$ in $G$, i.e.
 $\sigma$ is such that $Ad \; \sigma$ leaves $\h$ invariant and
$Ad \; \sigma|_{\h} = w$. There exists  $x \in \g$ (not unique) such
that $\sigma = \exp 2\pi ix$ and $ [x, \; \h_0] = 0$.
$\sigma $ is a finite oder inner automorphism of $\g$, $\sigma^m =1$,
than $\sigma$ is conjugate to an automorphism of $\g$ of the formed
$$
    \sigma' = exp\frac{2\pi i}{m} h, \; \; \; h \in \h(0)
$$
Recall some facts about twisted affine algebras (for more details see
\cite{K}). One can associate a subalgebra
$\hat{\got{L}}(\g,\, \sigma,\, m )$ of $\hg$ to the automorphism $\sigma$
of $\g$ as follows:
\be\label{S3}
 \got{L}(\g,\, \sigma,\, m ) =
\bigoplus \limits_{j \in \ZZ}   {\got{L}}(\g,\, \sigma,\, m)_j,
\ee
where ${\got{L}}(\g,\, \sigma,\, m)_j =\g_{j\:mod\: m}\tp t^j$ and
                                          $$
\g = \bigoplus \limits_{j \in \ZZ/m\ZZ} \g_j
$$
is the eigenspace decomposition for $\sigma$ and $\g_j$ is the eigenspace
of $\sigma$ for the eigenvalue $\ve^j$ ($\ve^j = \exp (2\pi i/m)$).
The decomposition (\ref{S3}) is a $\ZZ$-gradation of
${\got{L}}(\g,\, \sigma,\, m)$. We can extend this gradation on
\be\label{S4}
\hat{\got{L}}(\g,\, \sigma,\, m)= {\got{L}}(\g,\, \sigma,\, m)\oplus
\CC k \oplus \CC d,
\ee
setting $deg \,k = deg \, d = 0$. $\hat{\got{L}}_{\sigma}=
 \hat{\got{L}}(\g,\, \sigma,\, m)$ for arbitrary finite oder $\sigma$
is called a twisted affine algebra. If  $\sigma$ is an inner automorphism,
then \cite{K}
\be\label{S5}
  \hat{\got{L}}_{\sigma} \simeq \hg.
\ee
Let $\sigma$ be  an inner automorphism of $\g$, then we have  (\ref{S5}).
The twisted Heisenberg subalgebra corresponding to $w$ is
\be\label{S6}
\hat{\h}_w = \h_w(t) = \{ \h_{k \, mod\, m}
\tp t^k |\; \h_{k \, mod\, m} \in \g_k, \;
k \in \ZZ \},
\ee
where $\h_k$ is the centralizer of $\h_0$ in $\g_k$.

\subsection{Twisted evaluation modules and corresponding twisted\\ intertwining
operators}
We keep the notations $V(\Li), \; \; i \in I$ for the level one fundamental
modules of the simply laced Lie algebra $\hg$. The weights $P(V(\bLi))$
of $V(\bLi)$ are invariant under the action of the Weyl group $W$.
Fix $w \in W$ and consider the orbit decomposition under the action of $w$:
\be\label{o}
 P(V(\bLi)) = \bigsqcup^{n_i}_{j = 1} S_j.
\ee
Note that $n_i$ does not depend on $\bLi$ and is determined by $w$.
We will write $n_w$ instead of $n_i$.
Define $s_j = |S_j|$.

There exists a lifting $T$ of $w$ on $G$ satisfying the conditions
\be\label{O1}
T e_{\gamma} = e_{w(\gamma)},
\ee
for $\gamma \in P(V(\bLi))$ and a weight basis $\{e_{\gamma}\}$,
normalized as in (\ref{ba1}). Let $\gamma_j \in S_j$ be the maximal weight in
$S_j $ so, for $\beta \in S_j$, we have $\beta = \gamma_j - \al, \;  \;
\al \in Q_+$ and $\beta = w^k(\gamma_j)$ for some $k \in \ZZ$.
It is clear that $ V(\bLi)$ is a free  $\CC[T]$- module of rank $n_w$ over
the generators $ e_{\gamma_j}, \; j= 1, .. n_w$. Denote by $V_j$ the
linear span of $\{e_{\gamma}\}_{\gamma \in S_j}$
and by $T_j = T |_{V_j}$. We have $[ T_j, \; \h_0] =0$ for $j = 1, ..., n_w$
and $\h_0$ the stable point under the $w$ in $\h$. Then
\be\label{O2}
\h(0) = \h_0 \oplus ^{k_w}_{j = 1}\left (\sum_{l = 1}^{s_j} \CC T^l_j\right)
\ee
is a Cartan subalgebra in $\g$. We can choose the lifting $x$ of $w \in W$
in $G$, up to conjugacy classes, acting on $V(\bLi)$ as
\be\label{O4}
 x \cdot e_{w^k(\gamma_j)} = \ve_j^{-k}e_{w^k(\gamma_j)}, \; \; \;
\ve = e^{\frac{2\pi i}{s_j}}.
\ee
This determines the $m$-grading on $V(\bLi)$ consistent with the
$m$-grading on $\g$, given by (\ref{S3}),  with respect to the action
of m -order automorphism $x$:
\be\label{O6}
\g_j \cdot V_k \subset V_{k + j}.
\ee
Note that the orthogonal projection of
$\gamma^1, ... \gamma^s$ on $Q_w\tp \CC$ which we will denote by
$\gamma'^1, ... \gamma'^s$ form
 the basis $\Pi_w$ in the subspase  $Q_w\tp \CC$, stable under the action of
$w \in W$.
Due to (\ref{O2}) and (\ref{O4}) we can define the following basis in the
Heisenberg subalgebra $\hat{\h}_w$:
\be\label{H}
H^l_j  = T^l_j  \tp t^{\frac{ml}{s_j}}, \: \; l \neq 0\: mod \: s_j  \\
\gamma^i_k  = \gamma^i \tp t^{mk} , \; k \in \ZZ, \;\; i=1, ... ,s .
\ee
Consider the fields
\be\label{H1}
\gamma^i(z) = \sum_{k \in \ZZ} \gamma^i_k \, z^{-mk-1}, \\
H^j (z) = \sum_{l \neq 0\, mod \, s_j} H^l_j \, z^{- \frac{ml}{s_j} - 1}
\ee
and the corresponding formal  integrals over $z$ of these fields
(the Veneziano fields):
\be\label{H2}
\tilde{\gamma}^i(z) = i\sum_{k \in \ZZ} \gamma^i_k \,
\frac{z^{-mk}}{mk} - i \gamma^i - i\gamma ^i_0 \log z, \\
\tilde{H}^j (z) = i \sum_{l \neq s_j} H^l_j \,
\frac {z^{- \frac{ml}{s_j} }}{\frac{ml}{s_j}}.
\ee
These fields contain all generators of the Heisenberg subalgebra
$\hat{\h}_w$.
Define the twisted evaluation module
$$
V_t^w(\bLi) \;\simeq \; \bigoplus_{k \in \ZZ} (V_{k \: mod \, m} \tp t^k).
$$
Due to (\ref{O6}) the action of $\hat{\got{L}}_{\sigma}$ is compatible with
this gradation..

To each $w$-orbit $S_j \subset P(V(\bLi))$ one can associats a generating
function (field) on $V_t^w(\bLi)[z, z^{-1}]$:
\be\label{f}
E_j (z) = \sum_{k \in \ZZ}\; e_{w^k(\gamma_j)} \;t^{\frac{km}{s_j}}\;
z^{-\frac{km}{s_j}}.
\ee
For the adjoint twisted fundamental module
$$
(V_t^w(\bLi))^{*}  \simeq
\bigoplus_{k \in \ZZ} (V^{*}_{k \: mod \, m} \tp t^k).
$$
and the adjoint $w$- orbit $S^*_j$, one can associate an adjoint generating
function (field)
\be\label{f*}
E^{*}_j (z) = \sum_{k \in \ZZ}\; e^{*}_{w^k(\gamma_j)} \;t^{-\frac{km}{s_j}}\;
z^{\frac{km}{s_j}}.
\ee
It is known \cite{KP} that the realization (vertex-operator construction)
of the fundamental representation $V(\Li)$ are parametrised by the
element $w \in W$ (up to a conjugacy classes) in a sense
that the bosonization is based on the choice of the Heisenberg
subalgebra $\h_w(t)$ corresponding to  $w$ in the sense of (\ref{S6}).
We will denote this construction by $R_w$ and the corresponding
space of representation by  $V^{w}(\Li)$.

Consider the intertwining operator
 \be\label{I1}
\Phi_w^i(z) :  V^w(\Lambda_m)\longrightarrow
V^w(\Lambda_n)\tp (V^w_t(\bLi))^*[z,\, z^{-1}]\,
z^{(\bL_m|\bL_i)}
\ee
and the adjoint
 intertwining operator
\be\label{I2}
\Psi_w ^i(z) :  V^w(\Lambda_p)\longrightarrow
V^w(\Lambda_s)\tp V^w(\bLi)[z,\, z^{-1}] \,
z^{-(\bL_p|\bL_i)},
\ee
Each field $E_j(z)$  (\ref{f}) gives rise to the component of the
intertwining operator $\Phi_w^i(z)$
\be\label{fr}
\Phi_j^i(z) :  V^w(\Lambda_m)\longrightarrow
V^w(\Lambda_n)\tp \CC[[t]] \tp \CC[[z]]\, z^{<\bL_m | \bLi >}
\ee
and, correspondingly, the field $E^*_j(z)$ determined  from (\ref{I2}) the
 operator
\be\label{fr*}
\Psi_j^i(z) :  V^w(\Lambda_p)\longrightarrow
V^w(\Lambda_s)\tp \CC[[t]] \tp \CC[[z]]z^{- <\bL_p | \bLi >}.
\ee
The normalization by $z$ is determined by the common grading.
\subsection{Bosonization of twisted intertwining operators}

 Fix $w \in W$ and let $\hat{\h}_w$ be the corresponding
Heisenberg subalgebra (\ref{S6}).
Define the space of states az follows. Consider a sublattice $Q_w$ in $Q$
  stable under the action of $w$. Let $P_w$ be the lattice dual to $Q_w$.
Suppose we have a bimultiplicative cocycle $\ve_w(\al, \, \beta), \; \;
\al, \beta \in P_w$ on $P_w$ defined by the properties (\ref{coc1a})
-(\ref{coc1c}) for
the lattices $Q_w$ and $P_w$.
Consider the space
\be\label{X}
V_{P_w} = S_w \tp \CC_{\ve}[P_w],
\ee
where $S_w$ is a symmetric algebra over the space
\be
\h^{<0}_w = \sum_{n \in \ZZ/m \ZZ} \sum_{j<0} \h_{n }\tp t^{-n + jm}
\ee
and $\CC_{\ve}[P_w]$ is a twisted group algebra over the lattice $P_w$.
The space $V_{P_w}$ is a sum of all level one representations of
$\hg^w =  \got{L}(\g,\, \sigma,\, m ) $ in  the realization $R_w$.
This means that the bosonization of all operators is given in terms
of the Heisenberg subalgebra $\hat{\h}_w$ (\ref{S6}) and $\sigma$
is the lifting of $w$ to $Aut \g /Int \g$ determined as in (\ref{O4}).
Consider the basis in $\hat{\h}_w$ defined in (\ref {H}) and the
intertwining operators (\ref {fr}), (\ref{fr*}) defined by the
generating function (\ref{f}) on $V_t^w(\bLi)[z, z^{-1}]$ . From
the action (\ref{O1}) and the definition of the basis on $\hat{\h}_w$
one has the following commutation relations:
\be\label{It}
[H^l_p , \; \Phi_j^i(z)] =\; \delta_{p,j} \: z^{\frac{ml}{s_j}} \;
\Phi_j^i(z) ,
\ee
$$
 [\gamma^r_k, \; \Phi_j^i(z)] = \, <\gamma^r |\, \lambda_j> \;
\, z^{km} \; \Phi_j^i(z)
$$
and for the adjoint operators:
\be\label{In1}
[H^l_p, \; \Psi_j^i(z)] = - \delta_{p,j} \,\; z^{\frac{ml}{s_j}} \;
\Psi_j^i(z)
\ee
$$ [\gamma^r_k, \; \Psi_j^i(z)] =\, - <\gamma^r |\,\lambda_j> \,\; z^{km}
\; \Psi_j^i(z)
 $$
 This determines the operators $\Phi_j^i(z)$ and $\Psi_j^i(z)$
up to the action on $\CC_{\ve}[P_w]$. From the general commutation
relations we get the following expressions:
\be\label{In2}
\Phi_j^i(z) = \ve_w (-\al_j' , \bLi')
: e^{\frac{i}{s_j}{\tilde{H}^j (z)} + i\tilde{\lambda'}^j (z)}:
\ee
and
\be\label{In3}
\Psi_j^i(z) =(-1)^{(\al_j' , \bLi')} \ve_w (\al_j' , \al_j'- \bLi')
: e^{- \frac{i}{s_j} \tilde{H}^j (z) - i \tilde{\lambda'}^j (z)}:,
\ee
where $\al_j' , \; \bLi',\; \lambda' $  are the projections of the
  corresponding
weights on $Q_w \tp \CC$ parallel to the subspace $Q^{\perp}_w \tp \CC$ in
$P$ and $\lambda_j = \bLi - \al_j$ for $ \al_j \in \Delta_+$.
It is easy to see that this projection
is the same for all points of $S_j$.

We would like to point out, that the components of the operators
$\Phi_j^i(z)$, applied to the vacuum vector $|0>= 1 \tp 1 \in V_{P_w}$
give rise to the whole space $V_{P_w}$ and the fields
\be\label{aff}
X^i_{j,l, n}(z) = : \Phi_j^i(z) \: \Psi_l^i(\ve^n z):|_{V^w(\Lambda_m)}, \,
\; n = 1, ... , m
\ee
where $\ve = e^{\frac{2\pi i}{m} }$
define the  realization $R_w$ of the fundamental representation
$V(\Lambda_m)$ constructed in  \cite{KP}.
Due to (\ref{In2}) -(\ref{In3}) we get the
bosonization of the fields $X^i_{j,l}$ and their expression in terms
of the twisted intertwining operators.
The normal ordering is undestood
in a standard way:
\be\label{aff1}
\Phi_j^i(z) \: \Psi_l^i(w) = (z^m - w^m)^{ -(\lambda'^j, \lambda'^l)}
: \Phi_j^i(z) \: \Psi_l^i(w): \; \; {\rm for} \; \; |z| > |w|, \; l \neq j,
\ee
where $m$ is the order of $w \in W$.
\subsection{Hirota bilinear equation in terms of twisted\\ intertwining
operators}

We keep all notation of the above subsections.

Consider the operator
\be\label{B}
K^i_w(z) =\sum_{j = 1}^{n_w}  \Phi_{j}^i(z)
\tp
\Psi_{j}^i(z): \; \; V^w(\Lambda_m) \tp V^w(\Lambda_p) \longrightarrow
\ee
$$
 \ \:\; \; \; \: \; \; \; \; \;  \; \; V^w(\Lambda_n) \tp V^w(\Lambda_s)
\tp \CC[z,z^{-1}]\,z^{(\bLi' | \bL_m' - \bL_p')},
 $$
On $V^w(\Lambda_m) \tp V^w(\Lambda_p)$ operator
$K^i_w(z)$ has a decomposition
\be\label{R1}
K^i_w(z)= \sum_{k \in \ZZ }\,
K^i_k \, z^{-k + (\bLi' , \bL_m' - \bL_p')},
\ee
Where $\bLi'$ stands for
the projection to $Q_w \tp \CC$ parallel to  $Q^{\perp}$
(see (\ref{In2})-(\ref{In3})).

Define the $m$-vacuum $|m> = 1 \tp e^{\bL_m'}$ for the highest weight vector
in $V^w(\Lambda_m)$. For
\be\label{ku}
k > \max_{\gamma \in P_w(\Lambda_i)} < \bLi'- \gamma' |  \bL_m' - \bL_p'>
\ee

we have\\
1. $ K^i_{k} $, where $k =lm, \;l \in \ZZ$ and $m$  the order of $w \in W$,
commutes with the diagonal action of the Lie algebra $\hg^w$ on
$V^w(\Lambda_m) \tp V^w(\Lambda_p)$, given by (\ref{aff}).\\
2. $K^i_{k}  (|m> \tp |p>) = 0$ for $k$ as in (\ref{ku}).\\
3. The equation
\be\label{M}
 K^i_{lm} (\tau_1 \tp \tau_2) = 0, \; \; \;
\tau_1 \tp \tau_2 \in V^w(\Lambda_m) \tp V^w(\Lambda_p)
\ee
determines the hierarchy of soliton equations associated with the above
data and the vertex-operator construction $R_w$ of the
representations considered.

For  $\tau \in V^w(\Lambda_m)$ we have
$$
\tau = \sum_{\beta \in P_w(\Lambda_m)} \: \tau_{\beta}
\tp e^{ \beta},
$$
where $ \tau_{\beta} \in S_w$ and $P_w(V(\bL_m))$
(we will write  for simplicity $P_w(\bL_m)$) is the set of all
projections of $P(V(\bL_m))$ to $Q_w \tp \CC$ parallel to the $Q^{\perp}$.
We can identify the space $S_w$ with
the space $\CC [x^j_{k_j} , \; y^{\gamma}_l  ] $, where
$j = 1, ... , n_{w}, \; \; k_j \neq 0 \: mod \; s_j ; \; \;
\gamma \in \Pi_{w}$ ($\Pi_{w}$ is a basis in $Q_{\omega}$),
$l \in \ZZ$.

The action on $\CC [x^j_{k_j} , \; y^{\gamma}_l  ] $
of the Heisenberg subalgebra $\hat{\h}_w$ with the basis (\ref{H}) is
given by
\be\label{F}
\gamma_{-k} \to km\, y^{\gamma}_{k} ; \; \; \; \gamma_k \to \sum_{\beta \in
  \Pi_w} <\gamma | \; \beta > \frac{\partial}{\partial y^{\beta}_{k}} ; \; \;
\; \gamma_0 \to 0, \; \; \; k \in \ZZ_+
 \ee
 and
$$
H_l^j \to ml \, x^j_l, \;
\; H_{-l}^j \to \frac{\partial}{\partial x^j_l}, \; \;
l \in \ZZ_+ , \;\; l \neq 0\:mod \: s_j.
$$
On the space $\CC_{\ve} [P_w]$ the Heisenberg subalgebra $\hat{\h}_w$
acts by
$$
\gamma_k \cdot  e^{\beta} = \delta_{k, 0} \, < \gamma | \; \beta >\, e^{\beta};
\; \;
\; \; H_l^j \cdot e^{\beta} = 0,\; \; \; \beta \in P_w.
$$
By linearity for all $\beta \in  P_w$ $\beta_k$ is a well defined operator on
$V_{P_w}$. Similarly to the nontwisted case we will use the notation
$\partial_{\beta_l}$ for the operator $\beta _l$, $l > 0$.

Using the bosonisation formulas (\ref{In2}) - (\ref{In3}) for
$ \Phi_{j}^i(z)$ and $\Psi_{j}^i(z)$, one can write explicitly
the systems of Hirota bilinear equations. Fix $k = lm, \; l\in \ZZ$
satisfying (\ref{ku}) , then the equation (\ref{M}) turns into
\be\label{M1}
Res|_{z = 0} z^{lm - <\bLi' | \: \bL_m' - \bL_p' >} \left(
\sum_{\gamma_j \in
P_w(\bLi)} (-1)^{\bLi' | \, \bLi' - \gamma_j >} \ve_w(\bLi' - \gamma_j ,
-\bLi' - \gamma_j )\right.
\ee
$$
\left.: e^{\frac{i}{s_j}\tilde{H}^j (z) + i \tilde{\gamma}^j(z) } : \tp
: e^{- \frac{i}{s_j} \tilde{H}^j (z) - i \tilde{\gamma}^j(z) }: \right)
 \left(\sum_{\beta \in P_w(\bL_m), \delta \in  P_w(\bL_p)}\:
 \tau_{\beta} \, e^{ \beta}
 \tp \tau_{\delta} \, e^{ \delta} \right)=0.
 $$
 Using the action (\ref{F}) of the coefficiens of the fields $\tilde{H}^j (z)$
 and $\tilde{\gamma}^i(z)$ on $\CC [x^j_{k_j} , \; y^{\gamma}_l  ] $ and
 doing the standard manipulations as in (\ref{eq2}) - (\ref{eq4}),
 we can write the Hirota bilinear equations associated with the (i)-(iii)
 of the Definition 1.1.
 \begin{ex}
 Consider the case $\hg =\hsl (N, \CC)$.
 \end{ex}
 The Weyl group $W$ is isomorphic to the group $S_N$ of all $N$-point set
 permutations.  The conjugacy classes in $W$ are parametrised by the
 sequances $(n_1, ... , n_s ), \; n_1 \geq n_2, ... , \geq n_s$ and
 $\sum_{i = 1}^{s}n_i = N$.
 Fix $w \in W$ representing the class with the given sequance
 $(n_1, ... , n_s )$.
 As we have note (examlpe 2.1), all fundamental modules $V(\bLi), \;
 i \in I = \{0, 1, ... , N-1 \}$ are at level one. Fix $i\in I$ and consider
 the orbit decomposition (\ref{o}) under the the action of $w$ on the weight
 lattice:
 \be
 P(V(\bLi)) = \bigsqcup^{n_w}_{j = 1} S_j.
 \ee
 We have $n_w = s$ and $(s_1, ... , s_s)$ is
 a permutation of $(n_1, ... , n_s )$, where $s_j = |S_j|$.
 We can choose  the lifting  $x$ of $w \in W$ in $G$
   acting on $V(\bLi)$ as
\be\label{M4}
 x \cdot e_{w^k(\gamma_j)} = \ve_j^{-k}e_{w^k(\gamma_j)}, \; \; \;
\ve = e^{\frac{2\pi i}{s_j}},
\ee
where $\gamma_j$ is a maximal weight in $S_j$. The group element $x$
defines  compatible $m$-gradings on $\hg$ and on $V(\bLi)$
The orthogonal projections of $\gamma_j$ satisfy the relation
$\sum_{j=1}^s s_j \gamma_j^{'} =0$ and
$\{\gamma_j^{'}\}, \; j = 1, ... , s-1$ form a basis in $Q_w \tp \CC$.

 For more details, consider the case $\hg = \hsl(3, \CC)$. We have
 $W \simeq S_3$ and three possibilities of choice of nonisomorphic
 Heisenberg subalgebras, corresponding to the partitions
 $(1, 1, 1)$, $(2, 1)$, $(3)$. The first case corresponds to the
 homogeneous Heisenberg subalgebra and was cosidered in the second section.
 The thied case corresponds to the principal Heisenberg subalgebra.  This
 case can be obtained as a reduction from the one component KP and leads
 to the so-called Boussinesq hierarchy \cite{DJKM2}. The second case
 coresponds to the second order element $w \in W$. Let $\al_1, \al_2$ be the
 standard
 basis in $\hsl(3, \CC)$. Up to conjugacy class we can put $w = s_{\al_1}$
 the reflection corresponding to  $\al_1$. For the  action of $w$
 on the weights of $V(\bL_1)$ ($(V(\bL_2) = V^*(\bL_1)$) we have two orbits;
 \be
 S_1 = \{\gamma_1 = \frac{2}{3}\al_1 +\frac{1}{3}\al_2, \;
\gamma_2 = \gamma_1 - \al_1 =
 -\frac{1}{3}\al_1 +\frac{1}{3}\al_2 \} \; \; {\rm and} \;\;
 S_2 = \{\gamma_3 = -\frac{1}{3}\al_1 -\frac{2}{3}\al_2 \}.
 \ee
 The $\gamma_1^{'}= \frac{1}{6}\al_1 +\frac{1}{3}\al_2$ determines
 the one-dimensional space $Q_w \tp \CC$. The lattice $Q_w$
 is generated by $\al_2^{'} = \frac{1}{2} \al_1 + \al_2$ and
 the weight lattice $P_w$ by $\gamma_1^{'}=
 \frac{1}{6} \al_1 + \frac{1}{3}\al_2 $
  The grading element $x$ has
  the form
 $$
 x = diag \,(1, -1, 1).
 $$
 The basis in the graded Cartan subalgebra $\h_w$ is
 $$
 \gamma_1^{'} = \frac{1}{6}\al_1 +\frac{1}{3}\al_2\; \;{\rm and} \; \;
 T= X_{\al_1} + X_{-\al_1}.
 $$
 The basis in the corresponding Heisenberg subalgebra $\hat{\h}_w$ is
 $$
 \gamma_k =\gamma'_1 \tp t^{2k}, \; \;
  k \in \ZZ; \; \; \; \; H_l = T \tp t^l, \;
  \; l\neq 0 \,\: mod \: 2.
$$
We can identify the space
$S_w$ with the space $ \CC[x_1, x_3, ...; y_2, y_4, ...]$.
For the components (\ref {fr})-(\ref{fr*}) of the intertwining
 operators (\ref{I1})-(\ref{I2}) one has the following bosonization:
 \be\label{ru}

\Phi_1^1(z) = e^{\sum_{k \; odd, k\geq 0}
x_k z^k + \sum_{k \geq 0} y_{2k} \, z^{2k}}
e^{-\sum_{k \; odd, k\geq 0} \frac{z^{-k}}{2k}\frac{\partial}{\partial x_k}  -
\frac{1}{6}\sum_{ k\geq 0}
\frac{z^{-2k}}{2k}\frac{\partial}{\partial y_{2k}}} \tp e^{\gamma'_1}
(-z)^{\gamma_0}
\ee
$$
  \Phi_2^1(z) =
 e^{ -2\sum_{k \geq 0} y_{2k} \, z^{2k}}e^{\frac{1}{3}\sum_{ k\geq 0}
\frac{z^{-2k}}{2k}\frac{\partial}{\partial y_{2k}}} \tp e^{-2\gamma'_1}
(-z)^{-2\gamma_0}
$$
 and for the adjoint operators:
 \be
 \Psi_1^1(z) =  e^{-\sum_{k \; odd, k\geq 0}
x_k z^k - \sum_{k \geq 0} y_{2k} \, z^{2k}}
e^{\sum_{k \; odd, k\geq 0} \frac{z^{-k}}{2k}\frac{\partial}{\partial x_k}  +
\frac{1}{6}\sum_{ k\geq 0}
\frac{z^{-2k}}{2k}\frac{\partial}{\partial y_{2k}}} \tp e^{-\gamma'_1}
(-z)^{-\gamma_0}
\ee
$$
 \Psi_2^1(z) =
  e^{ 2\sum_{k \geq 0} y_{2k} \, z^{2k}}e^{- \frac{1}{3}\sum_{ k\geq 0}
\frac{z^{-2k}}{2k}\frac{\partial}{\partial y_{2k}}} \tp e^{2\gamma'_1}
(-z)^{2\gamma_0}.
$$
For $k=2l, \; \; l\geq 1$ we can write down the equation (\ref{M1})
in the following form. Identify the space
$$
V^w(\bL_m)
\simeq \CC[x'_n, \, y'_s] \tp \sum_{r \in \ZZ}e^{\frac{m}{3}\al'_2 + r \al'_2},
\; \; {\rm and} \;\;
V^w(\bL_p)
\simeq \CC[x^{''}_n, \, y^{''}_s]\tp \sum_{r \in \ZZ}e^{\frac{p}{3}\al'_2 + r
\al'_2},
$$
$n, \: s \in \ZZ, \; n$- odd, s - even. For $\tau^m \in V^w(\bL_m)$ we have
$$
\tau^m = \sum_{r \in \ZZ} \tau_{\frac{m}{3}+r }
\tp e^{\frac{m}{3}\al'_2 + r \al'_2}.
$$
Looking at
the coefficient of $e^{(\frac{m+1}{3} + r )\al'_2}
\tp e^{\frac{p-1}{3}\al'_2 + s \al'}$ in (\ref{M1}),
  $r, \: s \in \ZZ$, we get the following  equation:
\be\label{TT}
Res|_{z=0} z^{2l}
\left[
(-z)^{r-s}\right.
e^{\sum_{k \; odd, k\geq 0}
(x'_k -x^{''}_k)z^k + \sum_{k \geq 0} (y'_{2k} -y^{''}_{2k})\, z^{2k}} \cdot
\ee
$$
 e^{-\sum_{k \; odd, k\geq 0} \frac{z^{-k}}{2k}
(\frac{\partial}{\partial x'_k}-
\frac{\partial}{\partial x^{''}_k}) -
\frac{1}{6}\sum_{ k\geq 0}
\frac{z^{-2k}}{2k}(\frac{\partial}{\partial y'_{2k}} -
\frac{\partial}{\partial y^{''}_{2k}})} \tau_{\frac{m}{3}+r }(x',  y')
\tau_{\frac{p}{3}+s }(x^{''},  y^{''}) -
$$
$$
\left.
(-z)^{(p-m)+2(r-s-2)}
 e^{-2 \sum_{k \geq 0} (y'_{2k} -y^{''}_{2k})\, z^{2k}}
 e^{\frac{1}{3}\sum_{ k\geq 0}
\frac{z^{-2k}}{2k}(\frac{\partial}{\partial y^{'}_{2k}} -
\frac{\partial}{\partial y^{''}_{2k}})} \tau_{\frac{m}{3}+r+1 }(x', y')
\tau_{\frac{p}{3}+s -1}(x^{''},  y^{''})\right] =0.
$$
Introduce the new variables:
 $$
x_j = \frac{1}{2}(x_j' - x_j^{''}),\; \; \;
u_j = \frac{1}{2}(x_j' + x_j^{''})
$$
and
$$
y_j = \frac{1}{2}(y_j' - y_j^{''}),\; \; \;
v_j = \frac{1}{2}(y_j' + y_j^{''}).
$$
Thus
 $$
\frac{\partial}{\partial x_j} = \frac{\partial} {\partial
x_j^{'}} - \frac{\partial}{\partial x_j^{''}},\; \;
\frac{\partial}{\partial u_j} = \frac{\partial} {\partial
x_j^{'}} + \frac{\partial}{\partial x_j^{''}},\
 $$
 $$
\frac{\partial}{\partial y_j} = \frac{\partial} {\partial
y_j^{'}} - \frac{\partial}{\partial y_j^{''}},\; \;
 \frac{\partial}{\partial v_j} = \frac{\partial} {\partial y_j^{'}} +
\frac{\partial}{\partial y_j^{''}},\ $$
 Then the equation(\ref{TT}) is
 equivalent to the following system of Hirota bilinear equations:
 \be\label{TTT}

 \sum_{k \geq 0}p_k(2x, \, 2y)\, p_{k +2l+1+r-s}
 (-\frac{1}{2}\tilde{D}_x, -\frac{1}{6} \tilde{D}_y)
 e^{\sum_{p,l \geq1} x_{2p-1}D^x_{2p-1} +
 y_{2l}D^y_{2l}} \tau_{\frac{m}{3} +r}\, \tau_{\frac{p}{3} +s} -
 \ee
 $$
 (-1)^{p-m+r-s} \sum_{k \geq 0}p_k(-4 y)\,
 p_{k +2l+1+p-m +2(s-r-2)}(\frac{1}{3} \tilde{D}_y)
e^{\sum_{l \geq1}
 y_{2l}D^y_{2l}} \tau_{\frac{m}{3} +r+1}\, \tau_{\frac{p}{3} +s-1} =0,
 $$
where $p_s$ are the  Schur pollynomials depending on odd $x$ and even $y$
and $\tilde{D}_x$ stands for $(D_{x_1}, \frac{1}{3}D_{x_3}, ...)$,
$\tilde{D}_y$ stands for $(\frac{1}{2}D_{y_2}, \frac{1}{4}D_{y_4}, ...)$.
 Note that all these equations are invariant with respect to the
 translation
 $$
 m \to (m + k) \:mod \: 3, \; \; p \to (p + k) \:mod \: 3; \; \; \;
 r\to r + l, \; \; s \to s+l, \; \; k,\:l \in \ZZ.
 $$

 \subsection*{Acknowledgments}
The autors are grateful to A.Reyman for  useful comments and to
 C.Roger for hospitality during the stay in the Institute Girard
Desargues, URA 746, Universit\'{e} Lyon-1, where this work was completed.
 M.~G.-K is grateful to the Phys.Department of the University Roma 1 and
  the Institute des Hautes Etudes Scientifiques (France) for the kind
  hospitality and support during the preparation of this work.
The research of M.~G.-K and D.~L was supported in part
 by RFFI
grant 96-02-18046, grant  96-15-96455  for support of scientific schools 
and by Award No. RM2-150 of the U.S. Civilian Research \& Development
Fondation (CRDF) for the Independent States of the Former Soviet Union.


\begin{thebibliography}{*******}

\bibitem[S]{S} M. Sato, {\it Soliton equationa as dynamical systems on
infinit dimentional Grassmann manifold}, RIMS, {\bf 439}, (1981).
\bibitem[DJKM1]{DJKM1} E. Date, M. Jimbo, M. Kashiwara and T.Miwa
        {\it Operator approach to the Kadomsev-Petviashvili equation.
        Transformation group for soliton equations III}, J.Phys.Soc. Japan,
        {\bf 50}, (1981), 3806-3812.
\bibitem[DJKM2]{DJKM2} E. Date, M. Jimbo, M. Kashiwara and T.Miwa
{\it Transformation groups for soliton equations. Euclidean Lie algebras and
        reduction of the KP hierarchy}, Publ. Res. Inst. Math. Sci.
        {\bf 18}, (1982), 1077-1110.

\bibitem[K]{K}
  V.G. Kac { \it Infinite dimensional Lie Algebras},
  Third edition,
  Cambridge Univ. Press, (1990)

\bibitem[K1]{K1}
  V.G.~Kac,
  {\it Vertex algebras for beginners},
  University lecture series, ISSN, vol. {\bf 10},
  AMS, Providence RI, 1996.

\bibitem[KKLW]{KKLW}
  V.G. Kac, D.A. Kazhdan, J.~Lepowsky and R.L.~Wilson,
  {\it Realization of the basic representations of the Euclidean Lie
  algebras}, Adv. in Math,
  {\bf 4} (1981), 83--112.
\bibitem[KP]{KP}
  V.G. Kac and D.H. Peterson, {\it 112 constructions of the basic
        representation of the loop groop of $E_8$} in Proc. of
        the Symposium "Anomalies, Geometry, Topology", World
        Scientific, 1985, 276-289.
\bibitem[KV]{KV}
       V.G. Kac and J.van de Leur, {\it The n-Component KP Hierarchy
        and Representation Theory} in Important Develoopments in Soliton
        Theory, eds. A. Fokas and V. Zakharov. Springer Series in Nonlinear
        Dynamics, (1993), 302-343.
\bibitem[V]{V}
        J. van de Leur, {\it The $[n_1, n_2, ... , n_s]$-th Reduced KP
        Hierarchy and $W_{1+\infty}$ Constraine} in Proc. of the
        Int. Conf. on Math.Phys., String Theory and Quant. Gravity at
        Alushta, 1994.
\bibitem[GKK]{GKK} M.Golenishcheva-Kutuzova, V.Kac,
        $\Gamma$ -{\it Conformal algebras}, q-alg/9709006;
        J.Math.Phys,  vol 39, n.4, 2290-2305, 1998.
\bibitem[KW]{KW}
        V.G. Kac and M. Wakimoto, {\it Exeptional hierarchies of solyton
        equations}, Proc. Symposia in Pure Math, {\bf 141} (1989), 191-237.
\bibitem[FKRW]{FKRW}
        Ed. Frenkel, V. Kac, A. Radul and W. Wang {\it $W_{1+\infty}$
        and $W(\gl_N)$ with Central Charge $N$ }, Commun. Math. Phys, {\bf
        170}, (1995), 337-357.

\bibitem[FR]{FR}
        I.B. Frenkel and N.Yu. Reshetikhin, {\it Quantum Affine Algebras
        and Holonomic Difference Equations}, Commun. Math. Phys, {\bf 146},
        (1992), 1-60.
\bibitem[GKLMM]{GKLMM}
        A.Gerasimov, S.Khoroshkin, D.Lebedev, A.~ Mironov, A.~ Morozov,
        {\it Generalized Hirota Equations and Representation Theory
        The Case of $SL(2)$ and $SL_q(2)$}, Intern. J. of Modern Phys.,
        {\bf 10}, n. 18. (1995), 2589-2614.
\bibitem[KKL]{KKL} S.Kharchev, S.Khoroshkin, D.Lebedev, {\it
        Intertwining operators and Hirota bilinear equations},
        Theor.et Math.Phys.,{\bf 104}, n.1 (1995), 114-157.
        \bibitem[GK]{GK} M.
Golenishcheva-Kutuzova, {\it Soliton equations with Wakimoto Modules
        Symmetries. $\hslt$ - case}, in prepatation.
\bibitem[FZ]{FZ}
        V.A. Fateev and A.B. Zamolodchikov, {\it Parafermionic $Z_N$-models},
        Sov. J. Nucl. Phys., {\bf 43}, (1986).
\bibitem[W]{W} G.Wilson, Ergod.Th. and Dynam. Sys.1(1981)361;
Phil.Trans.R.Soc.Lond.  A 315(1985)383.
\bibitem[GHM]{GHM} M.F.de Groot, T.J. Hollowood and J.L.Miramontes,
Commun.Math.Phys. 145(1992)57.
\bibitem[BGHM]{BGHM} N.J.Burroughs, M.F.de Groot, T.J.Hollowood and J.L.
Miramontes, Commun.Math.Phys. 153(1993)187.
\bibitem[A]{A} H.Aratyn, L.A.Ferreira, J.F.Gomes and A.H.Zimerman,
J.M.P. 38(1997)1559;
Solitons from Dressing in an Algebraic Approach to the
Constrained KP Hierarchy, preprint solv-int/9709004.



\end{thebibliography}
\end{document}